\newif\ifsingle

\ifsingle
\documentclass[11pt,draftclsnofoot, onecolumn]{IEEEtran}		
\else		
\documentclass[10pt,final, twocolumn]{IEEEtran}
\fi

\usepackage{times}
\usepackage{amsmath,dsfont}
\usepackage{amssymb,amsthm}
\usepackage{epsfig,verbatim}
\usepackage{subcaption}
\usepackage{setspace}
\usepackage{color}
\usepackage{cite}
\usepackage{epstopdf}
\usepackage{graphics}
\usepackage{accents}
\usepackage{acronym}
\usepackage[bookmarks,colorlinks]{hyperref}
\usepackage{booktabs}
\usepackage{mathtools}
\usepackage{enumitem}
\usepackage{footmisc} 

\usepackage{algorithm}
\usepackage{algpseudocode}

\newcommand{\myVec}[1]{{\boldsymbol{#1}}}
\newcommand{\myMat}[1]{{\boldsymbol{#1}}}
\newcommand{\mySet}[1]{\mathcal{#1}}
\newcommand{\abs}[1]{\left| #1\right|}
\newcommand{\paren}[1]{\left( #1 \right)}
\newcommand{\cbrace}[1]{\left\{#1\right\}}
\newcommand{\sbrace}[1]{\left[#1\right]}
\newcommand{\E}{\mathds{E}}		 			
	
\newcommand{\argmin}[1]{\underset{#1}{\arg \min}}

\newcommand{\Nsparse}{k}
\newcommand{\Ninputs}{n}
\newcommand{\Noutputs}{m}
\newcommand{\Nlimit}{L}
\newcommand{\NlimitInTest}{l}
\newcommand{\Qset}{\mySet{Q}}

\newcommand{\pr}[1]{\Pr\paren{#1}}
\newcommand{\ex}[1]{\mathbb{E}\paren{#1}}

\newcommand{\off}[1]{}

\newcommand{\pt}{\tilde{p}}
\newcommand{\ptq}{\pt_{q}}
\newcommand{\ptu}{\pt_{u}}
\newcommand{\phu}{\hat{p}_{u}}
\newcommand{\ai}{A_{i}}
\newcommand{\aimu}{A_{i,j}}
\newcommand{\di}{D_{i}}
\newcommand{\mmu}{M_{j}}
\newcommand{\wmu}{W_{j}}

\newcommand{\AcD}{\myMat{A}_{\cD}}
\newcommand{\rank}[1]{\text{rank}\paren{#1}}
\newcommand{\reals}{\mySet{R}}
\newcommand{\qzovy}{Q_{0,1}\paren{\vy}}
\newcommand{\qzovz}{Q_{0,1}\paren{\vz}}

\newtheorem{theorem}{Theorem}

\newtheorem{proposition}{Proposition}
\newtheorem{lemma}{Lemma}

\definecolor{NewColor}{rgb}{0,0,0} 

\definecolor{DarkGreen}{rgb}{0.1,0.5,0.1}
\definecolor{DarkRed}{rgb}{0.5,0.1,0.1}
\definecolor{DarkBlue}{rgb}{0.1,0.1,0.5}
\definecolor{DarkPurple}{rgb}{0.5,0.2,0.5}
\definecolor{DarkTurquoise}{rgb}{0.1,0.5,0.5}

\newcommand{\cD}{{\cal D}}

\newcommand{\cK}{{\cal K}}

\newcommand{\cN}{{\cal N}}
\newcommand{\cP}{{\cal P}}
\newcommand{\cS}{{\cal S}}

\newcommand{\cW}{{\cal W}}

\newcommand{\vu}{\myVec{u}}
\newcommand{\vv}{\myVec{v}}

\newcommand{\vx}{\myVec{x}}
\newcommand{\vy}{\myVec{y}}
\newcommand{\vz}{\myVec{z}}

\newcommand{\bigo}[1]{\mathcal{O}\paren{#1}}

\newcommand{\rev}[1]{{\color{black}{#1}}}

\ifsingle

\else

\fi 

\acrodef{adc}[ADC]{analog-to-digital convertor}
\acrodef{cdc}[CDC]{Centers for Disease Control and Prevention}
\acrodef{cs}[CS]{compressed sensing}
\acrodef{gt}[GT]{group testing}
\acrodef{dtft}[DTFT]{discrete-time Fourier transform}
\acrodef{dnn}[DNN]{deep neural network}
\acrodef{csi}[CSI]{channel state information}
\acrodef{map}[MAP]{maximum a-posteriori probability}
\acrodef{snr}[SNR]{signal-to-noise ratio}
\acrodef{bs}[BS]{base station}
\acrodef{iot}[IOT]{Interent of Things}
\acrodef{mimo}[MIMO]{multiple-input multiple-output}
\acrodef{mse}[MSE]{mean-squared error}
\acrodef{pdf}[PDF]{probability density function}
\acrodef{rv}[RV]{random variable}
\acrodef{ml}[ML]{maximum likelihood}
\acrodef{ls}[LS]{least squares}
\acrodef{fec}[FEC]{forward error correction}
\acrodef{rs}[RS]{Reed-Solomon}
\acrodef{lti}[LTI]{linear time-invariant}
\acrodef{dnd}[DND]{definitely not defective}
\acrodef{dd}[DD]{definitely defective}
\acrodef{pd}[PD]{possible defective}
\acrodef{psd}[PSD]{power spectral density}
\acrodef{ser}[SER]{symbol error rate}
\acrodef{ber}[BER]{bit error rate}
\acrodef{sgd}[SGD]{stochastic gradient descent}
\acrodef{isi}[ISI]{intersymbol interference}
\acrodef{awgn}[AWGN]{additive white Gaussian noise}
\acrodef{ut}[UT]{user terminal}
\acrodef{mmw}[mmWave]{millimeter wave}
\acrodef{noma}[NOMA]{non-orthognal multiple access}
\acrodef{mac}[MAC]{mulitple access channel}
\acrodef{fl}[FL]{Federated learning}
\acrodef{pcr}[RT-qPCR]{qualitative reverse transcription polymerase chain reaction}
\acrodef{omp}[OMP]{Orthogonal Matching Pursuit }
\acrodef{rna}[RNA]{Ribonucleic acid}

\IEEEoverridecommandlockouts

\title{One-Shot Pooled COVID-19 Tests via\\ Multi-Level Group Testing}

	\author{
		\IEEEauthorblockN{Amit Solomon, Alejandro Cohen, Nir Shlezinger, Yonina C. Eldar, and Muriel M\'edard
		}
			\thanks{
		Parts of this work were presented at the 2021 IEEE International Conference on Acoustics, Speech, and Signal Processing as the paper \cite{cohen2021ICASSP}.	
		This project received funding from the Miel de Botton and Jean and Terry de Gunzburg Coronavirus Research, by the Manya Igel Centre for Biomedical Engineering and Signal Processing, by the QuantERA grant C’MON-QSENS!, by the European Union’s Horizon 2020 research and innovation program under grant No. 646804-ERC-COG-BNYQ, and from the Israel Science Foundation under grant No. 0100101.
		 A. Solomon, A. Cohen, and M. M\'edard are with the	Research Laboratory of Electronics, MIT, Cambridge, MA (email: \{amitsol, cohenale, medard\}@mit.edu).
		N. Shlezinger  is with the School of ECE, Ben-Gurion University of the Negev, Be'er-Sheva, Israel  (e-mail: nirshl@bgu.ac.il).
		Y. C. Eldar is with the Faculty of Math and CS, Weizmann Institute of Science, Rehovot, Israel (e-mail: yonina@weizmann.ac.il).
		}
	\vspace{-0.75cm}
	}

\begin{document}
	
	\maketitle
	\pagestyle{plain}
	\thispagestyle{plain}
	
\begin{abstract}
	A key requirement in containing contagious diseases, such as the Coronavirus disease 2019 (COVID-19) pandemic, is the ability to efficiently carry out mass diagnosis over large populations. Some of the leading testing procedures, such as those utilizing qualitative polymerase chain reaction, involve using dedicated machinery which can simultaneously process a limited amount of samples. A candidate method to increase the test throughput is to examine pooled samples comprised of a mixture of samples from different patients. In this work we study pooling-based tests which operate in a one-shot fashion, while providing an indication not solely on the presence of infection, but also on its level, without additional pool tests, as often required in  COVID-19 testing. As these requirements  limit the application of traditional group-testing (GT) methods, we propose a multi-level GT scheme, which builds upon GT principles to enable accurate recovery using much fewer tests than patients, while operating in a one-shot manner and providing multi-level indications. We provide a theoretical analysis of the proposed scheme and characterize conditions under which the algorithm operates reliably and at affordable computational complexity. Our numerical results demonstrate that multi-level GT  accurately and efficiently detects infection levels, while achieving improved performance over previously proposed one-shot COVID-19 pooled-testing methods.
	
\end{abstract}

\vspace{-0.5cm}
\section{Introduction}\label{sec:intro}
\vspace{-0.1cm}
The Coronavirus disease 2019 (COVID-19) pandemic has already forced lockdowns all over the globe, and has claimed more than three million lives worldwide.
In order to handle and contain pandemics, and particularly COVID-19, large portions of the population should be frequently tested \cite{salathe2020covid}.
One of the main difficulties in doing so stems from the limited testing resources and the lengthy duration required to identify the presence of an infection \cite{emanuel2020fair}.
In particular, the main diagnosis tool for COVID-19 tests is based on \ac{rna} extraction via \ac{pcr}. Although various technological alternatives have been proposed \cite{lucia2020ultrasensitive,ben2020sars}, the \ac{pcr} process remains the leading 
method for COVID-19 testing. The output of this procedure represents an estimate of the viral load in the tested sample \cite{nolan2006quantification}. The main bottleneck associated with this form of COVID-19 testing follows from the fact that each \ac{pcr} machine can simultaneously process a fixed number of samples, and its procedure tends to be quite lengthy, on the order of a few hours for each test.

A promising method to tackle \rev{this lengthy measurement} procedure and thus to increase the efficiency of COVID-19 tests is based on pooling \cite{yelin2020evaluation, hanel2020boosting}. Here, each sample processed, i.e., each input to the \ac{pcr} machine, is comprised of a mixture of several samples taken from different patients. When the infected patients constitute a small subset of the overall tested individuals, pooling-based schemes allow accurate recovery using a number of tests which is notably smaller than the number of patients \cite{gilbert2008group}. Several recovery schemes for pooled COVID-19 tests have been recently proposed
\cite{hanel2020boosting,ben2020large, shental2020efficient ,ghosh2020compressed,yi2020low,petersen2020practical, yi2020error, zhu2020noisy}, which can be divided according to the two main mathematical frameworks for such recovery procedures: The first is \ac{gt} theory, originally derived for detecting infections in large populations \cite{dorfman1943detection}, used in \cite{yelin2020evaluation, hanel2020boosting, ben2020large}. \ac{gt} \rev{was} first suggested to identify syphilis-infected draftees during World War II~\cite{dorfman1943detection}, and has been long studied and utilized in many fields, such as biology and chemistry~\cite{du2000combinatorial,macula1999probabilistic}, communications~\cite{varanasi1995group,cheraghchi2012graph,wu2014partition}, sensor networks~\cite{bajwa2007joint}, pattern matching~\cite{clifford2007k}, web services~\cite{tsai2004testing}, \rev{and} cyber security~\cite{cormode2005s,xuan2009detecting,goodrich2005indexing}\rev{.} The second framework is \ac{cs}, which deals with the recovery of sparse signals \cite{eldar2012compressed}, and was utilized for pooled COVID-19 tests in \cite{shental2020efficient,ghosh2020compressed, yi2020low,petersen2020practical, yi2020error, zhu2020noisy}.

One of the main differences between classical \ac{gt} and \ac{cs} is that \ac{gt} deals with group detection problems, which results in binary variables. Specifically, in \ac{gt} each subject can either be infected or not infected \cite{gilbert2008group}, while \ac{cs}-based methods result in real-valued variables. \ac{gt} is traditionally adaptive, requiring multiple sequential tests\cite{dorfman1943detection} in order to achieve a minimal number of overall tests from which the presence of infection can be inferred. Nonetheless, \ac{gt} can also be applied in a one-shot (non-adaptive) manner \cite{chan2014non}, avoiding the need to mix new samples during the testing procedure. \ac{cs} focuses on the one-shot recovery of sparse real-valued vectors from lower-dimensional linear projections, and thus each subject can take any real value number \cite{eldar2012compressed}. The additional domain knowledge of \ac{gt}, namely, the more restricted binary domain over which it operates compared to \ac{cs}, allows it in some applications to \rev{operate using fewer} measurements \rev{compared to \ac{cs}}, as \rev{shown} in \cite{cohen2019serial,cohen2020distributed} \rev{in the context of} quantization of sparse signals.

\rev{When} testing for contagious diseases, and particularly for COVID-19, one is often interested \rev{in obtaining} some score on the level of the viral load \rev{of the patients due to its epidemiological value \cite{ghosh2020compressed,beldomenico2020superspreaders,liu2020viral}}\rev{. This can be achieved} using \ac{cs} tools. The fact that \rev{\ac{gt} and \ac{cs} have their own} \off{each of these mathematical frameworks has its own} pros and cons for pooled testing, motivates the design of a recovery method which combines \ac{gt} with one-shot operation and multi-level detection, as in \ac{cs}.

In this work we propose a multi-level \ac{gt} recovery scheme for pooled testing. Our proposed \ac{gt}-based method is designed to account for the unique characteristics of pooled tests for contagious diseases, and particularly those arising in COVID-19 \rev{testing}. The proposed technique extends \ac{gt} schemes to detect multiple levels of viral load, building upon our previous results on combining \ac{gt} with \ac{cs} concepts and multi-level discretization in \cite{cohen2019serial}. The resulting multi-level \ac{gt} scheme operates in a one-shot manner, and is designed to avoid dilution due to mixing too many samples in each pool \cite{yi2020optimal}.

We begin by identifying the specific requirements which arise from the setup of pooled COVID-19 testing. \off{In light of} \rev{From} these requirements, we derive the multi-level \ac{gt} method. Our scheme  is comprised of $a)$ a dedicated testing matrix, determining which patients are pooled together into \rev{each test}; and $b)$ a \ac{gt}-based recovery method operating in a two stage manner, by first identifying the defective patients building upon classic \ac{gt} \rev{tools}, followed by a dedicated mechanism for characterizing the level of infection for \rev{the}\off{those} identified patients.

We theoretically analyze the proposed multi-level \ac{gt} scheme. We first characterize its associated computational complexity, which is shown to be dominated by the identification of the defective subjects in the first stage of the proposed algorithm. As the complexity formulation results in a random quantity which depends on the statistical modelling of the measurement procedure, we derive the expected computational burden, and show that it results in a number of computations which is of the same order as that of low-complexity \ac{cs}-based methods.  Next, we derive sufficient conditions for the algorithm to yield a unique solution.
While similar guarantees are also available for \ac{cs}-based methods, we numerically demonstrate that our proposed   scheme achieves improved accuracy over \ac{cs}-based pooled recovery. Our experimental results use the model proposed in \cite{ghosh2020compressed} for pooled \ac{pcr} testing. For \rev{these} setups, we demonstrate that \rev{our}  multi-level \ac{gt} scheme reliably recovers the infection levels, while operating at a limited computational burden, and \rev{achieves} improved accuracy over existing \ac{cs}-based approaches.

The rest of this paper is organized as follows:  Section~\ref{sec:Model} reviews the system model, focusing on pooled COVID-19 testing  and identifies the unique requirements of this procedure. Section~\ref{sec:MultiGT} presents the proposed multi-level \ac{gt} scheme. In Section~\ref{sec:pd_calc} \rev{we analyze our approach}, identifying sufficient conditions for it to reliably detect the level of infection of each patient, and \rev{characterize} its complexity.
Section~\ref{sec:sims} details the simulation study, and Section~\ref{sec:Conclusions} provides concluding remarks.
	
Throughout the paper, we use boldface lower-case letters for vectors, e.g., ${\myVec{x}}$.
Matrices are denoted with boldface upper-case letters,  e.g.,
$\myMat{M}$. Let $\myMat{M}\paren{i,j}$ denote the element at the $i$-th row and $j$-th column of $\myMat{M}$.
Sets are expressed with calligraphic letters, e.g., $\mySet{X}$, and $\mySet{X}^n$ is the \rev{$n$-th} order Cartesian power of $\mySet{X}$.
The stochastic expectation is denoted by   $\E\{ \cdot \}$,  $ \bigvee$ and $\oplus$ are the Boolean OR and XOR operations, respectively, 
\rev{and} $\mySet{R}_{+}$ is the set of non-negative real numbers. 

\vspace{-0.1cm}
\section{System Model}\label{sec:Model}
In this section we present the system model for which we derive the recovery algorithm in Section~\ref{sec:MultiGT}. We begin by identifying the specific characteristics of pooled COVID-19 tests in Subsection~\ref{subsec:Model_Assumptions}, based on which we present our problem formulation in Subsection~\ref{subsec:Model_Problem}.


\begin{figure*}
	\centering
	\includegraphics[width=0.8\linewidth]{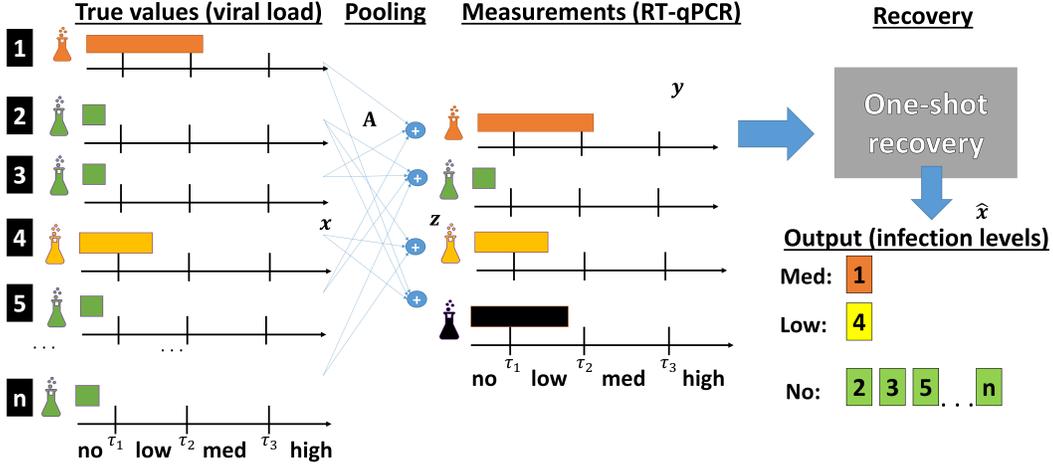}
	\caption{Pooled \ac{pcr} testing with one-shot recovery.\vspace{-0.5cm}}
	\label{fig:LearningTech1}
\end{figure*}

\vspace{-0.2cm}
\subsection{Pooled COVID-19 Tests}
\label{subsec:Model_Assumptions}
\vspace{-0.1cm}
The common approach in testing for the presence of COVID-19 is based on the \ac{pcr} method. Here, a sample is collected, most commonly based on nasopharyngeal or oropharyngeal swabs or saliva. The presence of infection is then examined by \ac{rna} extraction via \ac{pcr} measurements, quantifying the viral load in the sample. The \ac{pcr} process is quite time consuming (on the typical order of several hours), and can simultaneously examine up to a given number of $\Noutputs$ inputs (commonly on the order of several tens of samples). This results in a major bottleneck, particularly when the number of patients, denoted by $\Ninputs$, is much larger than $\Noutputs$.

A candidate approach to reduce the test duration, which is considered in this paper, utilizes pooling \cite{yelin2020evaluation}. Pooling mixes the samples of groups of patients together, forming $\Noutputs$ mixed samples out of the overall $\Ninputs$ patients. Then, the presence of COVID-19 for each of the tested individuals is recovered from the mixed \ac{pcr} measurements, either directly, i.e., in a {\em one-shot} fashion, or in an adaptive manner involving additional tests \cite{hughes1994two}. \rev{Some characteristics of COVID-19 tests are:}
\begin{enumerate}[label={\em A\arabic*}]
	\item\hspace{-0.13cm}: \label{itm:sparsity} The number of infected measurements, denoted by $\Nsparse$, is much smaller than the number of tested individuals $\Ninputs$. Typically up to $10\%$ of the tested population is infected.
	\item\hspace{-0.13cm}: \label{itm:Val} One is  interested not in only identifying whether a subject is infected, but also in some discrete score on the viral load. For example, possible outputs are {\em no} (no virus), {\em low} (borderline), {\em mid} (infected), and {\em high} (highly infected).
	\item\hspace{-0.13cm}:  \label{itm:Noise} The  \ac{pcr} measurements are noisy, i.e., some level of random distortion is induced in the overall process, encapsulating the randomness in the acquisition of the samples, their mixing, and the  \ac{pcr} reading. 
	\item\hspace{-0.13cm}: \label{itm:OneShot} One-shot tests \rev{are preferable, where we} fully identify all subjects from a single \ac{pcr} operation, without having to carry out additional tests based on the results.
	\item \hspace{-0.13cm}:\label{itm:Limit} There is a limit, denoted by $\Nlimit > 1$, on the number of subjects which can be mixed together in a single measurement. A typical limit on the number of subjects mixed together is $\Nlimit = 32$ \cite{yelin2020evaluation}. Furthermore, the portion taken from each sample for the pooled measurements is identical, e.g., one cannot mix $30\%$ from one patient with $70\%$ from another patient into a single pool.
\end{enumerate}
While the characteristics are identified for swab-based \ac{pcr} tests, they also hold for other forms of contagious disease testing, including, e.g., serological tests \cite{Allen2021}.

An illustration of the overall flow of pooled \ac{pcr}-based COVID-19 testing along with the desired one-shot recovery operation is depicted in Fig. \ref{fig:LearningTech1}. On the left side of the figure, we \rev{show} the true viral loads of all $n$ items., where the first item is infected in a medium level, the fourth item is infected in a low level, and all other items are not infected. Next, pooling is done based on a testing matrix, which is generated prior to obtaining the samples. For example, the first pooled test involves samples from the first, third, and fifth items. This results in a \rev{measurement} vector, denoted by $\myVec{z}$. This vector is fed to the recovery algorithm, which is able to tell in one-shot that the first item is infected in a medium level, the fourth item is infected in a low level, and all other items are not infected.

\vspace{-0.3cm}
\subsection{Problem Formulation}
\label{subsec:Model_Problem}
\vspace{-0.1cm}
Based on the characteristics of pooled COVID-19 tests detailed above, we consider the following problem:
Let $\myVec{x} \in \mySet{R}_{+}^{\Ninputs}$ be a vector whose entries are the viral loads of the $\Ninputs$ patients. By \ref{itm:sparsity} it holds that  $\myVec{x}$ is $\Nsparse$-sparse, i.e., $\|\myVec{x}\|_0 \leq \Nsparse$.
The pooling operation is represented by the matrix $\myMat{A} \in \{0,1\}^{\Noutputs\times \Ninputs}$.
 Let $\NlimitInTest (i) \leq \Nlimit$  denote the number of subjects mixed together in the \rev{$i$-th} individual pool, $i\in\{1,\ldots,m\}$. This implies that the \rev{$i$-th} row of $\myMat{A}$, denoted $\myMat{A}_i^T$, is $\NlimitInTest(i)$-sparse by \ref{itm:Limit}. The viral loads of the pooled samples are represented by the vector $\myVec{z} \in \mySet{R}_{+}^{\Noutputs}$, whose entries are given by
\begin{equation}
\label{eqn:PooledViral}
\myVec{z}_i = {\frac{1}{\NlimitInTest(i)}}\myMat{A}_i^T\myVec{x}, \quad i\in\{1,\ldots,m\},
\end{equation}
where the factor $\frac{1}{\NlimitInTest(i)}$ and the structure of $\myMat{A}$ guarantee that identical portions are taken from each sample in a pool-test, as required in  \ref{itm:Limit}.
The \ac{pcr} measurements, denoted by $\myVec{y} \in \mySet{R}_{+}^{\Noutputs}$, are given by some stochastic mapping $f: \mySet{R}_{+}\mapsto  \mySet{R}_{+}$ applied  to $\myVec{z}$. This mapping represents the distortion detailed in \ref{itm:Noise}, and we write the measurements as $\myVec{y} = f(\myVec{z})$, where $f(\cdot)$ is carried out element-wise.

To formulate our objective, we note that by \ref{itm:Val}, we are interested in recovering a discrete representation of the viral load. We thus define the discretization mapping $Q:\mySet{R}_{+}\mapsto \Qset$, where $\Qset$ is a finite set containing the possible decisions, e.g., $\Qset = \{no, low, med, high\}$. The decision regions are defined by the \rev{pre-specified} thresholds $\tau_1,\ldots, \tau_{|\Qset|-1}$, such that a value not larger than $\tau_1$ is treated and not infected, while, e.g., a value in the range $(\tau_1,\tau_2]$ is treated as a low level of infection.
Our goal is thus to design an algorithm which maps the \ac{pcr} measurements $\myVec{y}$ into an estimate of the discretized viral loads, denoted by $\hat{\myVec{x}}\in\mySet{Q}^{\Ninputs}$\rev{. We wish to minimize} the error probability, defined as
\begin{equation}
\label{eqn:error}
e(\hat{x}) \triangleq \frac{1}{\Ninputs}\sum_{i=1}^{\Ninputs} \Pr\left(Q(x_i) \neq \hat{x}_i \right).
\end{equation}
The fact that  $\hat{\myVec{x}}$ is obtained directly from $\myVec{y}$ indicates that the algorithm operates in a one-shot fashion, as required in \ref{itm:OneShot}. By \ref{itm:sparsity}, we have prior knowledge of a reliable upper bound on the number of infected patients $\Nsparse$. \rev{Such a bound is often obtained by pooling asymptomatic subjects, for which the infection ratio is typically low, on the order of $1\%$ \cite{shental2020efficient}.} When $\Nsparse$ is not known or the given bound is expected to be loose, one of the methods proposed in the \ac{gt} literature to approximate this number within $\bigo{\log n}$ tests can be used,  e.g.,~\cite{damaschke2010competitive,damaschke2010bounds}.

To summarize, for the subset of $\Nsparse$ infected items of a total of $\Ninputs$ inspected patients, the goal in multi-level pooled testing is to design an $\Noutputs\times \Ninputs$ one-shot pooling pattern $\myMat{A}$ and a corresponding recovery algorithm for reconstructing $\hat{\myVec{x}}$ from $\myVec{y}$. The measurement matrix should guarantee that at most $\Nlimit$ subjects are mixed in each pool-test. \rev{We aim to} identify the subset of infected items and their discrete representation of the viral load using the recovery algorithm \off{, from the observed vector $\myVec{y}$.}\rev{from $\myVec{y}$.}

\vspace{-0.1cm}
\section{Multi-Level Group Testing}\label{sec:MultiGT}
\vspace{-0.1cm}
In this section, we provide an efficient scheme which implements \ac{gt} with multiple decisions. Our design builds upon the fact that the sparsity assumption \ref{itm:sparsity} implies that the recovery of pooled \ac{pcr} tests can be treated as a sparse recovery problem,
which is typically studied under either the framework of \ac{gt} \cite{gilbert2008group}, or that of \ac{cs} \cite{eldar2012compressed}. Broadly speaking, \ac{gt} deals with sparse recovery of binary variables, i.e., it can recover whether a subject is infected or not. In order to evaluate the actual levels of each tested subject, as requested in \ref{itm:Val}, one would have to re-run the \ac{pcr} test, violating requirement \ref{itm:OneShot}. The alternative approach of \ac{cs} operates over the domain of real numbers, namely, it attempts to identify the exact cycle threshold or viral load for each subject, and thus tends to be less accurate compared to \ac{gt}, as it does not exploit the fact that one is only interested in a discrete grade value by \ref{itm:Val}. This motivates the derivation of a dedicated algorithm for pooled COVID-19 recovery, which harnesses the strength of \ac{gt} theory while extending it to the multi-level domain. 
The proposed multi-level \ac{gt} method is presented in Subsection~\ref{subsec:proposed}, followed by a  discussion in Subsection~\ref{subsec:Discussion}.

\vspace{-0.3cm}
\subsection{Pooled-Testing Algorithm}\label{subsec:proposed}
\vspace{-0.1cm}
Multi-level \ac{gt} is comprised of two components: The design of the testing matrix $\myMat{A}$, which determines the pooling operation; and the recovery algorithm which determines the discrete level associated with each subject based on the results of the pooled tests. We next elaborate on each of these components.
\off{ An example is given in Fig.~\ref{fig:LearningTech1}. On the left side of the figure, we see the true viral loads of all $n$ items. In particular, we see that the first item is infected in a medium level, the fourth item is infected in a low level, and all other items are not infected. Next, pooling is done based on previously generated testing matrix. For example, the first pooled test involves samples from the first, third, and fifth items. This results in a \rev{measurement} vector, denoted by $\myVec{z}$. This vector is fed to the recovery algorithm, which is able to tell that the first item is infected in a medium level, the fourth item is infected in a low level, and all other items are not infected.}

\subsubsection{Pooling Procedure}\label{subsec:testing_matrix}
To determine \off{the testing matrix} $\myMat{A}$, we first fix the number of the pool-tests $\Noutputs$. In principle, $\Noutputs$ should satisfy $\Noutputs = (1+\epsilon)\Nsparse\log_2\Ninputs$, for some $\epsilon>0$, as this is the sufficient number of pool-test for reliable recovery in \ac{gt} using the optimal \ac{ml} decoder \cite{atia2012boolean,cohen2016secure}. \rev{The parameter $\epsilon$ controls the probability of error of the procedure \cite{atia2012boolean}, as we illustrate in Section \ref{sec:sims}.} In practice, the number of pools-tests is often dictated by the measurement setup, e.g., it may be constrained to be an integer multiple of the number of inputs accepted by an \ac{pcr} machine. Unless stated otherwise, we assume that $\Noutputs = (1+\epsilon)\Nsparse\log_2\Ninputs$, and in Section~\ref{sec:pd_calc} we prove that this number of samples is sufficient to detect the infection levels using the \rev{proposed} algorithm.

Once $\Noutputs$ is fixed, we proceed to setting $\myMat{A}$, whose binary entries dictate which patient is mixed into which pool.
The traditional \ac{gt} method of generating $\myMat{A}$ draws its elements in an i.i.d. fashion according to a Bernoulli distribution with parameter $p$. A common choice for $p$ is $p=1-2^{-1/k}$, for which the probability of each element in $\myVec{z}$ to be zero is $1/2$. \rev{When $k$ is unknown,} $p$ is chosen using a rough approximation of $k$.
A major drawback of this approach is that~\ref{itm:Limit} is not necessarily satisfied, and there is some chance \rev{that} too many patients will be mixed into the same pool causing dilution. We therefore consider an alternative method, which forces the columns of $\myMat{A}$, as well as the rows of $\myMat{A}$, to be ``typical". That is, we want every column/row to have exactly $\lceil p\cdot m \rceil \leq \Nlimit$ and $\lceil p\cdot n\rceil$ ones, respectively. \rev{This ensures that with high probability half of the elements of $\myVec{z}$ are zero, which in turn reduces the required pooled tests, as demonstrated in Section \ref{sec:sims}. This requirement cannot be guaranteed in the non-asymptotic regime by generating the elements of $\myMat{A}$ in an i.i.d. fashion, hence we force $\myMat{A}$ to be typical.}

Since in practical testing setups, one is interested in using a fixed deterministic matrix, rather than having to work with random matrices, we generate $\myMat{A}$ once before the pooling starts. \rev{A typical matrix is not unique, and in fact there are many typical matrices. Generating a typical matrix, that satisfies \ref{itm:Limit} and the number of required pool tests for recovery, can be done readily.} 
\rev{The} same testing matrix can be used for multiple pooling experiments.


\subsubsection{Recovery Algorithm}\label{subsec:recovery_algorithm}
The proposed recovery algorithm is given in Algorithm~\ref{alg:recovery}. It operates in two main steps\rev{:} The first step treats the recovery as a \ac{gt} setup, and thus aims to divide the patients into infected and non-infected. The second step accounts for the multi-level requirement \ref{itm:Val}, and uses  decisions made in the first step to \rev{determine} the   infection levels.

{\bf Step 1: Detection of Defective Subjects.}
The first part of the algorithm identifies efficiently all of the \ac{dd} items in two stages, without determining the infection levels. It does so by treating the problem, in which the observations and the viral loads take continuous non-negative values in general, as a \ac{gt} setup which operates over binary variables. Hence, in this step we use a binary representation of the real-valued $\vy$, denoted  $\qzovy$, where $Q(\cdot)_{0,1}$ divides each measured pool into infected and non-infected. Recalling that the viral level decisions are defined by the thresholds $\{\tau_1,\ldots, \tau_{|\Qset|-1}\}$, the \rev{$i$-th} element of $\qzovy$, denoted $[\qzovy]_i$, is given by
\begin{equation}
    [\qzovy]_i =\begin{cases}
    0 & y_i \leq \tau_1, \\
    1 & y_i >\tau_1,
    \end{cases} \qquad i=1,\ldots, m.
\end{equation}

\begin{algorithm}\small
\caption{Multi-level \ac{gt} recovery}
\label{alg:recovery}
 \vspace{-1.5mm}
\begin{flushleft}
Input: $\myMat{A}, \myVec{y}$\newline
Output: $\cS$ \Comment{$\cS$ contains index and infection level tuples}
 \vspace{-1.5mm}
\end{flushleft}
\begin{algorithmic}[1]
\State $\cS\gets\emptyset$
\Statex\textbf{\underline{Step 1}: Detection of defective subjects}
\State $\cP\gets\text{DND}(Q_{0,1}(\myVec{y}),\myMat{A})$ \Comment{$\cP$ contains PD items}\label{line:DND}
\State $\cD\gets\text{ML}(Q_{0,1}(\myVec{y}),\myMat{A},\cP)$ \Comment{$\cD$ contains DD items}\label{line:ML}
\State $\myMat{\hat{A}}\gets \myMat{A_{\cD}}$
\Statex\textbf{\underline{Step 2}: Infection level recovery}
\vspace{0.4mm}
\Statex\textbf{\underline{Option 1}: Least squares method}
\State Solve \ac{ls} problem \eqref{eqn:LS}\label{line:least_squares}
\ForAll{$i$ s.t. $i\in \cbrace{1,\ldots,\abs{\cD}}$}
\State $\cS\gets\cS\cup\cbrace{\cD\paren{i},Q\paren{\vx_{\cD_{i}}}}$\label{line:quantize_least_squares}
\EndFor
\Statex\textbf{\underline{Option 2}: Iterative method}
\vspace{0.4mm}
\State $\bar{\myVec{y}}\gets\sum_{i\in\cD}\myMat{A}_i$\label{line:infection_detection_start}
\While{$|\cS|<K$} \label{line:while_S_k}\label{line:while_not_done}
\ForAll{$i$ s.t. $\bar{\myVec{y}}_i=1$} \label{line:sum_one}
\State $j\gets$ index s.t. $\myMat{\hat{A}}_{i,j}=1$\label{line:find_one}
\State $\hat{\myVec{x}}_i\gets Q(\myVec{y}_i)$\label{line:quantize}
\State $\cS = \cS\cup \left\{ (\cD(j), \myVec{\hat{x}}_i) \right\}$\label{line:save}
\State $\bar{\myVec{y}}\gets \bar{\myVec{y}} - \myMat{A}_j$\label{line:subtract_viral_load}
\State $\vy\gets \vy - \myMat{A}_j \cdot \vy_i$\label{line_subtract_viral_load_real}
\EndFor
\EndWhile\label{line:infection_detection_end}
 \vspace{-1.5mm}
\Statex\hrulefill
\Statex Definitely Not Defective (\ac{dnd}) \cite{chan2014non}
\vspace{-1.5mm}
\Statex\hrulefill
\Procedure{DND}{$Q_{0,1}(\myVec{y}),\myMat{A}$}\label{line:DND_start}
\State $\cP\gets\left\{1,\ldots,n \right\}$
\ForAll{$i$ s.t. $Q_{0,1}(\myVec{y})_i=0$}
\ForAll{$j$ s.t. $\myMat{A}_{i,j}=1$}
\State $\cP\gets\cP\setminus\left\{ j\right\}$
\EndFor
\EndFor
\State \Return $\cP$
\EndProcedure\label{line:DND_end}
 \vspace{-1.5mm}
\Statex\hrulefill
\Statex Maximum Likelihood (\ac{ml}) \cite{atia2012boolean} over \ac{pd} subjects $\cP$
\vspace{-1.5mm}
\Statex\hrulefill
\Procedure{ML}{$Q_{0,1}(\myVec{y}),\myMat{A},\cP$}\label{line:ML_start}
\State \Return $\arg\min_{\tilde{\cD} \subset \Omega(\cP,k)}\qzovy \oplus \left(\bigvee_{k\in\tilde{\cD}} \myVec{A}_k \right)$
\EndProcedure\label{line:ML_end}
\end{algorithmic}
\end{algorithm}

In the first stage of the first step, the \ac{dnd} algorithm \cite{aldridge2014group} is used (lines~\ref{line:DND} and \ref{line:DND_start}-\ref{line:DND_end}). Recall that the number of pool-tests $\Noutputs$, dictated by the testing matrix $\myMat{A}$, is fixed to allow \ac{ml} detection. The \ac{dnd} algorithm attempts to match the columns of $\myMat{A}$ with the vector $\qzovy$. In particular, if column $j$ of $\myMat{A}$ has a non-zero entry while the corresponding element in $\qzovy$ is zero, the column is assumed not to correspond to a defective subject. This algorithm finds most of the subjects that are \ac{dnd} and drastically reduces the number of \ac{pd} items. The set of subjects declared as \ac{pd} after this stage, denoted \rev{by} $\cP$, is shown to be indeed much smaller than the number of patients $n$; this is proved rigorously in Section~\ref{sec:pd_calc}, and demonstrated empirically in our numerical study in Section~\ref{sec:sims}. In particular, in Section~\ref{sec:pd_calc} we give a precise expression \rev{for} the expected size of $\cP$. We show that $\ex{\abs{\cP}}\ll n$ in the non-asymptotic regime that is of interest for COVID-19 pooled testing, and numerically assert these claims, showing that typically $|\cP| < 3\Nsparse$.  The remaining $n-|\cP|$ subjects are declared not defective.

The fact that the number of \ac{pd} subjects is notably smaller than the overall number of patients is exploited in the second stage of the first step, which determines the set of \ac{dd} patients.
 Here, the Boolean \ac{ml} algorithm \cite{atia2012boolean} is used only over the smaller set of \ac{pd} subjects $\cP$,  to identify exactly the set $\cD$ of $\Nsparse$ \ac{dd} subjects (lines \ref{line:ML} and \ref{line:ML_start}-\ref{line:ML_end}). We use \ac{ml} over the subset $\cP$ rather than over all the patients, allowing us to carry \rev{out} this typically computationally prohibitive method at affordable complexity, as we show in Subsection~\ref{subsec:complexity}. The \ac{ml} algorithm looks for a collection of $\Nsparse$ columns in $\myMat{A}$, for which $\qzovy$ is most likely. \rev{The \ac{ml} decision rule} is given by:
 \begin{equation}
 \label{eqn:GTML}
 \cD = \argmin{\tilde{\cD} \subset \Omega(\cP,k)} \qzovy \oplus \left(\bigvee_{k\in\tilde{\cD}} \myVec{A}_k \right).
 \end{equation}
 In the \ac{ml} rule we denote by $\cK$ the set of defective subjects, and by $\Omega(\cP,k)$ the set of $\binom{|\cP|}{k}$ combinations of $k$ defective subjects in $\cP$. While the formulation here is given with the true number of infected patients $\Nsparse$, in practice it is carried out with $\Nsparse$ (which is unknown) replaced by an approximation or an available upper bound on it.

{\bf Step 2: Infection Level Recovery.} The output of the  first step is a division of the patients into infected and non-infected, encapsulated in the set of identified defective subjects $\cD$. In the second step, the algorithm estimates the infection level of each subject in  $\cD$. We provide two  methods to do so:

{\em Option 1 - least squares:} The first option uses \ac{ls} estimation ~\cite{dekking2005modern} to identify the viral loads of all \ac{dd} items (line~\ref{line:least_squares}):
\begin{equation}
\label{eqn:LS}
\hat{\myVec{x}}_{\cD}= \argmin{\myVec{x}_{\cD}} \|\myVec{y} - \myMat{A}_{\cD}\myVec{x}_{\cD}\|_2^2,
\end{equation}
where $\AcD$ denotes the matrix created by taking the columns of $\cD$ from $\myMat{A}$.
The output of the algorithm is the infected items, and the quantized infection levels $Q\paren{\vx_{\cD}}$ (line~\ref{line:quantize_least_squares}) using the threshold-based quantization mapping $Q(\cdot)$ defined in Subsection~\ref{subsec:Model_Problem}.
\rev{For the \ac{ls} solution in \eqref{eqn:LS} to be unique, we should have $\rank{\AcD}=k$. In this case,}  \eqref{eqn:LS} is given by
\begin{equation}
\label{eqn:LSsol}
\hat{\myVec{x}}_{\cD}=\paren{\myMat{A}_{\cD}^T\myMat{A}_{\cD}}^{-1}\myMat{A}_{\cD}^T\myVec{y}.
\end{equation}

\rev{When the measurement process induces some known non-linearity, it can be incorporated into \eqref{eqn:LS}, resulting in a non-linear \ac{ls} formulation.} The \ac{ls} operation outputs a real-valued vector, while the discrete levels are obtained by quantizing the entries of \eqref{eqn:LSsol}. \off{\rev{Note that $\myVec{x}_{\cD}$ has to be non-negative. This constraint is satisfied by the unconstrained \ac{ls} when the solution is unique in the scenario considered by this paper. To see this, consider \eqref{eqn:LSsol}. Since both the pooling matrix $\myMat{A}$ and the measurement vector $\myVec{y}$ are non-negative, the right hand side of \eqref{eqn:LSsol} is non-negative, thus the unique solution is non-negative. In Subsection~\ref{subsec:correctnext_least_squares} we prove that the probability that a unique solution exists is high.}}
\off{\rev{Note that $\myVec{x}_{\cD}$ has to be non-negative. This constraint is satisfied by the unconstrained \ac{ls} in the scenario considered by this paper. To see this, recall that a solution to \eqref{eqn:LS} satisfies
\begin{equation}\label{eqn:LSsol_any}
    \myMat{A}^T\myMat{A}\myVec{x}=\myMat{A}^T\myVec{y}.
\end{equation}
Both the pooling matrix $\myMat{A}$ and the measurement vector $\myVec{y}$ are non-negative, hence the right hand side of \eqref{eqn:LSsol_any} is non-negative. This forces $\myVec{x}$ to be non-negative.}}
\rev{Note that $\myVec{x}_{\cD}$ has to be non-negative, thus one can compute the \ac{ls} in \eqref{eqn:LS} while constraining the entries of  $\myVec{x}_{\cD}$ to be non-negative. Here, we omit this constraint; The simulations conducted in Section~\ref{sec:sims} show that omitting this constraint does not affect the performance of the algorithm in the case studied in this paper. However, in the general setting this constraint can be easily included if needed \cite{lawson1995solving}.} 
\rev{We} to choose this method in our analyses \rev{below}, namely, the complexity analysis and the theoretical performance guarantees characterized in the sequel consider Algorithm~\ref{alg:recovery} implemented with the \ac{ls} recovery option. Nonetheless, we also provide an additional method to recover the infection levels based on iterative detection.

{\em Option 2 - iterative detection:} An alternative approach is to iteratively estimate the viral loads using pools containing a single patient whose viral load was not yet recovered, and to update the measurements accordingly. \rev{In contrast to option 1, which jointly maps the entries of $\myVec{y}$ into an estimate of the viral loads, here we iteratively search for the pools which represent a single unknown value, identifying their corresponding infection levels separately, and then canceling their contribution on the remaining pools.}  This approach facilitates recovery in setups where not only the infection levels, but also the viral loads themselves are discrete.

The iterative method for recovering the infection levels from $\cD$ is summarized in lines \ref{line:infection_detection_start}-\ref{line:infection_detection_end} of Algorithm~\ref{alg:recovery}. Here, we let $\cD(i)$ be the $i$-th element of $\cD$. For a test in which only one infected subject participates according to the testing matrix (lines \ref{line:sum_one}-\ref{line:find_one}), the algorithm can recover the viral load directly from the measurement (lines \ref{line:quantize}-\ref{line:save}). To obtain a discrete score, the measured value is quantized  using a threshold-based  mapping $Q(\cdot)$. Then the algorithm subtracts the viral load of that subject from all the tests in which it participates (lines \ref{line:subtract_viral_load}-\ref{line_subtract_viral_load_real}), and repeats until it recovers the infection levels of all declared infected subjects (line \ref{line:while_not_done}).

In Section~\ref{sec:sims} we compare the performance of the algorithm with \ac{ls} method vis-à-vis the iterative method. In our numerical results, which focus on a scenario with continuous-valued viral loads, the \ac{ls} method is shown to achieve more accurate recovery. This follows from its ability to mitigate the effects of the noise induced in the measurement process, allowing  it to retrieve viral loads that are close enough to the true viral loads. \rev{Option 2, in which detected pools are iteratively subtracted from non-detected ones, is more sensitive to noise and inaccurate identifications made in previous iterations, whose error may propagate to subsequent decisions.}

\vspace{-0.3cm}
\subsection{Discussion}
\label{subsec:Discussion}
\vspace{-0.1cm}
The novelty of the multi-level \ac{gt} algorithm stems from its two-step procedure; the first step utilizes efficient \ac{gt} methods to notably reduce the number of examined patients, while the second step inspects the remaining patients and applies dedicated mechanisms to recover their level of infection. Here, we note a few remarks arising from the proposed scheme.

In Subsection~\ref{subsec:proposed} we describe how the  matrix $\myMat{A}$ is generated. The description involves  random generation, for which the resulting matrix is not guaranteed to satisfy \ref{itm:Limit}. The motivation for using such randomization stems from its provable performance guarantees in \ac{gt} theory \cite{gilbert2008group}. \rev{In practice, }once a typical testing matrix satisfying \ref{itm:Limit} is selected, one does not have to generate a new matrix for each group of $\Ninputs$ patients.

\rev{For $\Ninputs$ i.i.d. tested individuals, the probability of finding the infected items (though not necessarily their levels) is maximized by the \ac{ml} rule}. However, its complexity is burdensome, as it has to consider $\binom{n}{k}$ options~\cite{coja2020information,aldridge2019group}. An efficient alternative is the \ac{dnd} algorithm, also known as column matching~\cite{kautz1964nonrandom,chan2014non,aldridge2014group}, whose time complexity is $\bigo{kn\log n}$~\cite{cohen2016secure}. However, it requires a greater amount of pooled measurements compared to \ac{ml} in order to reliably identify the detective items. The first step of our proposed multi-level \ac{gt} method combines the concepts of \ac{dnd} with the \ac{ml} algorithm, while the second step extends them to operate over non-binary fields, i.e., recover multiple levels rather than just identifying which subject is defective. Performing \ac{dnd} on all $\Ninputs$ items using the number of tests set to allow \ac{ml} detection, i.e., $\Noutputs = (1+\epsilon) \Nsparse \log_2 \Ninputs$ \rev{as opposed to the number of tests required for \ac{dnd}, which is $e\paren{1+\epsilon}k\log_{2}n$ \cite{chan2014non}, results in a smaller set of \ac{pd} subjects $\cP$. This in turn notably reduces the complexity of the recovery method}. In Section~\ref{sec:pd_calc} we give exact expressions for $\ex{\abs{\cP}}$ after performing \ac{dnd}. Given $\cP$, the \ac{ml} decoder has to consider significantly less options, $\binom{|\cP|}{k}$, which is likely to be computationally feasible and considerably faster than considering all $\binom{n}{k}$ combinations, as discussed in the complexity analysis in Subsection~\ref{subsec:complexity}. 

Algorithm~\ref{alg:recovery} requires two conditions to hold. First, the number of \ac{pd} items, i.e., $\abs{\cP}$, should be relatively small, preferably close to $\Nsparse$\rev{, so that the \ac{ml} algorithm is feasible}. Furthermore, for the \ac{ls} recovery method to have a unique solution, it should hold that $\rank{\AcD}=k$. \rev{If the solution is not unique, the algorithm may produce an undesirable output of different decisions when applied to the same pooled measurements.} Both these requirements are numerically asserted to hold in the experimental scenarios considered in the paper, and their validity is also theoretically studied in the following section. 

\rev{We note that in this paper, we assume that the binary quantization of the tests $\qzovy$ does not have any error. If this does not hold, one can use the noisy variants of the \ac{dnd} and the \ac{ml} algorithms of Step 1 as known in the \ac{gt} literature \cite{atia2012boolean,chan2014non}. We leave this case for future work.} 

\rev{In} our numerical evaluations of multi-level \ac{gt} we demonstrate its ability to achieve error-free detection when the number of measurements $m$ satisfies the {\em upper} bound on this quantity for which the computationally prohibitive \ac{ml} \ac{gt} is guaranteed to approach zero error, i.e., $m =\paren{1+\epsilon}k\log_2 n$ for some $\epsilon>0$~\cite{atia2012boolean}. Furthermore, in the case where it is possible to identify the {\em number} of infected subjects in a pool-test by observing its measurement, i.e., when one can separately tell for each element of $\myVec{y}$ how many defective patients are mixed into it (though not necessarily their infection level), we notice empirically that it is possible to achieve the {\em lower} bound number of pool test suggested for \ac{gt} using \ac{ml} algorithm, i.e., $m \approx \log_2 \binom{\Ninputs}{\Nsparse} \leq \Nsparse \log_2 (\Ninputs/\Nsparse) $ \cite{atia2012boolean,chan2014non} (for further \rev{details} see Section~\ref{sec:sims}). We leave the analysis of this case for future work.

\vspace{-0.3cm}
\section{Theoretical analysis}\label{sec:pd_calc}
\vspace{-0.1cm}
In this section, we provide a theoretical analysis of the proposed multi-level \ac{gt} scheme. In particular, we characterize two key components of Algorithm~\ref{alg:recovery}: the expected number of \ac{pd} subjects, which dominates the computational complexity of the algorithm, and guarantees for \ac{ls} recovery to have a unique solution.
We then proceed to analyze the complexity of Algorithm~\ref{alg:recovery}, and show it is computationally efficient in the regimes of \rev{interest} in the context of COVID-19. Our theoretical results are given as \rev{a} function of the system parameters, i.e., the overall number of patients $n$ and the number of infected ones $k$ (assumed in this section to be the true number of infected patients and not a bound on it), as well as using those of the pooling pattern, namely, the number of measurements $m$ and the Bernoulli distribution parameter $p$. While our theoretical measures are given using the true number of infected patients, we recall that Algorithm~\ref{alg:recovery} can be applied using an approximation or an upper bound on this quantity. For the analysis in this section, we assume that $\myMat{A}$ is drawn according to an \rev{i.i.d.} Bernoulli distribution with parameter $p$. 

\rev{In} our analysis of the expected number of \ac{pd} subjects, \rev{we consider different models of the noise in} 
Step 1. Following conventional \ac{gt} terminology \cite{atia2012boolean} \rev{the cases we treat are:}
\begin{itemize}
    \item {\em Noiseless case} - here the noise induced by the measurement process does not affect the ability to identify whether or not a pool contains a defected subject. Namely, $\qzovy = \qzovz$.
    \item {\em Additive noise} - here the presence of noise may result in a pool not containing any defected subject being measured as infected pool. This implies that $Q_{0,1}(\myVec{y}_i) \geq Q_{0,1}(\myVec{z}_i)$ for each $i \in 1,\ldots, \Noutputs$.
    \item {\em Dilution noise} - dilution implies that pools containing infected subjects may be falsely measured as non-defective. In this case, $Q_{0,1}(\myVec{y}_i) \leq Q_{0,1}(\myVec{z}_i)$ for each $i \in 1,\ldots, \Noutputs$.
\end{itemize}
These cases are studied individually in Subsections-\ref{subsec:Noiseless}-\ref{subsec:dilution}, respectively, for arbitrary \rev{settings} of the parameters $k,n,p,m$; conditions for uniqueness of the \ac{ls} solution are stated in  Subsection~\ref{subsec:correctnext_least_squares}, and the complexity \rev{of} Algorithm~\ref{alg:recovery} is characterized in Subsection~\ref{subsec:complexity}.

\vspace{-0.2cm}
\subsection{Noiseless Case}
\label{subsec:Noiseless}
\vspace{-0.1cm}
We first calculate how many items are declared \ac{pd} \rev{after running \ac{dnd} in} Algorithm~\ref{alg:recovery} in the noiseless setting. Recall that in this stage we use the \ac{dnd} algorithm to declare the \ac{pd} subjects. In the noiseless setting, all defective $k$ items are declared \ac{pd}, and there is some probability that each non-defective item is declared \ac{pd} by the \ac{dnd} algorithm. For Algorithm~\ref{alg:recovery} to run in reasonable time, it is essential that $\abs{\cP}$ is small enough, so the number of options to be considered by the \ac{ml} decoder $\binom{\abs{\cP}}{k}$ is not exponentially large. In the noiseless setting, the expected number of \ac{pd} subjects $\ex{\abs{\cP}}$ is stated in the following theorem:
\begin{theorem}\label{thm:noiseless}
The expected number of items declared \ac{pd} by \ac{dnd} algorithm (first stage of the first step of Algorithm~\ref{alg:recovery}) in the noiseless setting is given by
\begin{eqnarray}
&&\ex{\abs{\cP}} = k+\paren{n-k}\paren{1-p\paren{1-p}^k}^m. \label{eq:ex_pd_noiseless}
\end{eqnarray}
\end{theorem}

\begin{IEEEproof}
The proof is given in Appendix \ref{app:Proof1}.
\end{IEEEproof}

Theorem~\ref{thm:noiseless} characterizes the exact expected number of \ac{pd} items declared by \ac{dnd} in the noiseless setting. This characterization, which does not appear in the \ac{gt} literature, holds regardless of the fact that the outputs of the \ac{dnd} method are later used for identifying infection levels. Note that the second part of~\eqref{eq:ex_pd_noiseless} is similar to a known result \rev{on} the \ac{dnd} algorithm, e.g.~\cite[Eq. (8)]{chan2011non}. The main difference is that it is traditionally used to bound the probability of error of the algorithm using \rev{a} union bound, whereas we use this expression to calculate the exact expected number of \ac{pd} items.

\begin{figure}
    \centering
    \includegraphics[width=1 \columnwidth]{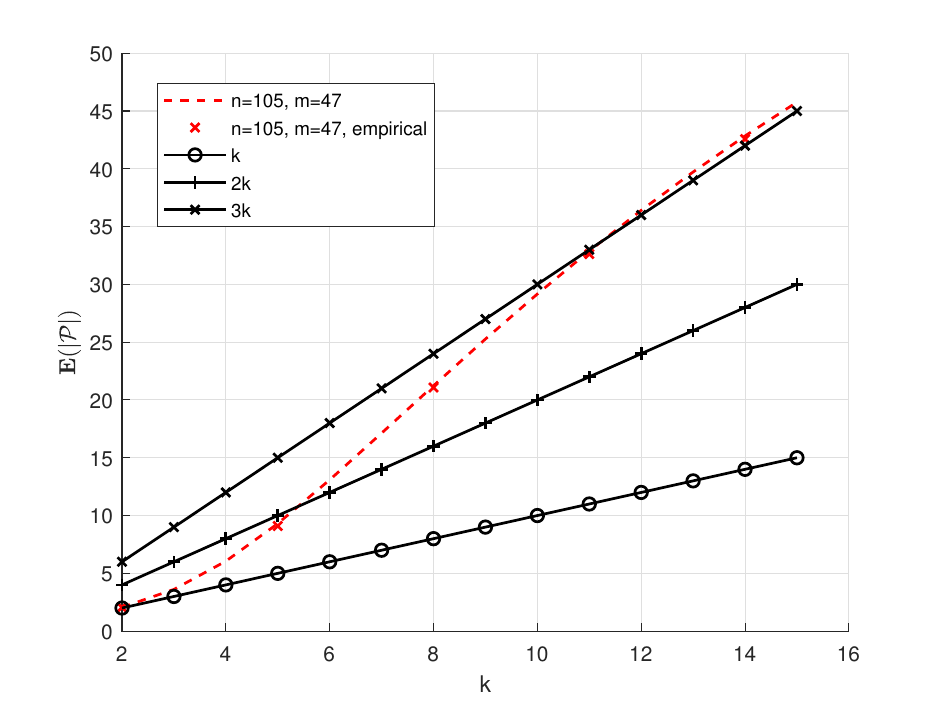}\vspace{-0.2cm}
    \caption{Simulated and theoretical number of elements declared PD by DND in the noiseless case,  $k=5$, $n=105$, and $m=47$, matching the ML upper bound (See Section~\ref{sec:sims}).}
    \label{fig:noiseless_sim}
\end{figure}

\begin{figure}
    \centering
    \includegraphics[width=1 \columnwidth]{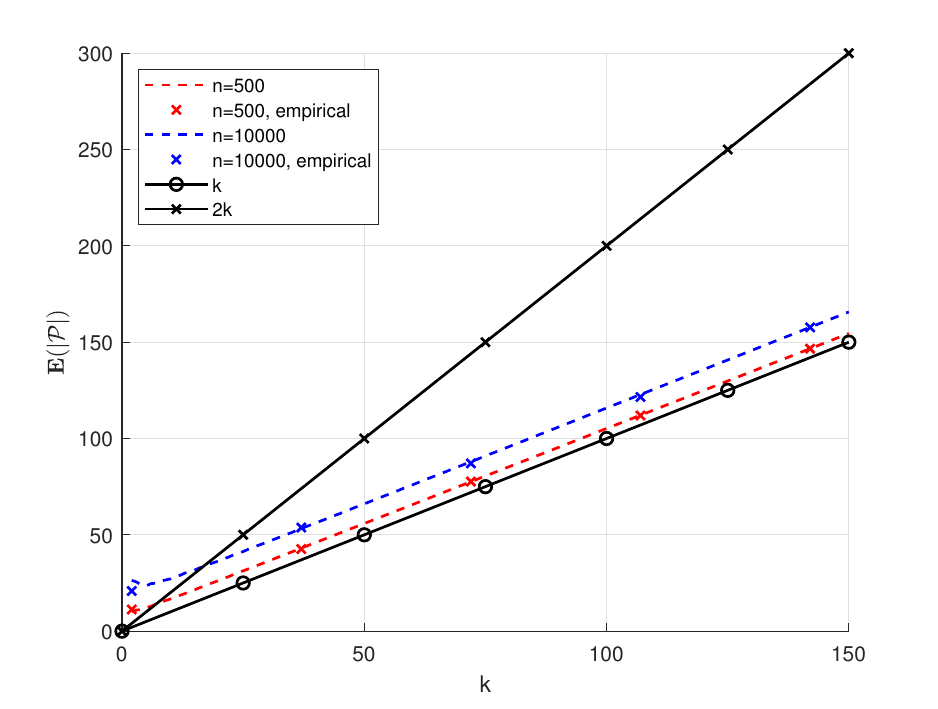}\vspace{-0.2cm}
    \caption{Simulated and theoretical number of elements declared PD by DND in the noiseless case, for a high number of $n$. The number of tests $m$ satisfies $m=1.4k\log_2 n$.}
    \label{fig:noiseless_sim2}
\end{figure}

To assert and numerically evaluate Theorem~\ref{thm:noiseless}, we depict in Fig.~\ref{fig:noiseless_sim} the expected number of items declared \ac{pd}, computed via \eqref{eq:ex_pd_noiseless}, plotted as the dashed lines, as a function of the number of defective items $k$. This theoretical quantity is compared to the empirical expected number of items declared \ac{pd}, averaged over $10^{4}$  Monte-Carlo simulations. The number of tests are chosen to match the upper \ac{ml} bound $m=\paren{1+\epsilon}k\log_2 n$, as described in Section~\ref{sec:sims}. We observe in Fig.~\ref{fig:noiseless_sim} that the simulation results agree with the theoretical values dictated by Theorem~\ref{thm:noiseless}. The figure also shows that in the non-asymptotic regime, which is of interest for applications such as COVID-19 pooled testing, the expected number of \ac{pd} items is significantly smaller than the number of patients $n$, and in particular, often lies in $\sbrace{k,3k}$. This means that the second part of Step 1 of Algorithm~\ref{alg:recovery}, which uses an \ac{ml} decoder, has to consider much less combinations compared to that needed without the prior \ac{dnd} subjects identification. Similar observations are reported in Fig.~\ref{fig:noiseless_sim2} which considers the same comparison as in Fig.~\ref{fig:noiseless_sim}, only with a significantly larger number of items $n$.
The figure shows that the average number of \ac{pd} items after \ac{dnd} is substantially smaller than $n$, indicating that the gains of using \ac{dnd} in reducing the computational burden are more dominant as the number of patients grows.

\vspace{-0.3cm}
\subsection{Additive Noise}\label{subsec:Noise}
\vspace{-0.1cm}
In the additive noise model, a test may \rev{be} positive even if no defective items participate in it~\cite{atia2012boolean}. This case reflects the possibility of false positives in traditional binary \ac{gt}. This happens with probability $q$, i.i.d. between different tests. That is, we define a set of i.i.d. random variables $\cbrace{W_i}_{i=1}^m$, such that $W_i$ obeys a Bernoulli distribution with parameter $q$.
The binarized test results of the $i$-th test is here modelled as a Boolean OR function of the test result without additive noise and $W_i$ i.e.,
\begin{equation}
    Q_{0,1}(\myVec{y}_{i}) = Q_{0,1}(\myVec{z}_{i})\bigvee W_i, \quad i \in 1,\ldots, m.
\end{equation}
It follows from~\cite{atia2012boolean} that the \ac{dnd} algorithm can still guarantee accurate recovery of the \ac{pd} subjects\off{. However, for this guarantee to hold,}\rev{ when} the number of tests \off{should be}\rev{is} increased by a factor of at least $1/\paren{1-q}$ with respect to the noiseless scenario. We expand the results of Theorem~\ref{thm:noiseless} in the following theorem:

\begin{theorem}\label{thm:additive_noise}
Let $q$ be the probability that a test without infected items comes out positive. The expected number of items declared \ac{pd} by \ac{dnd} algorithm (first stage of the first step of Algorithm~\ref{alg:recovery}) in the additive noise setting is given by
\begin{eqnarray}
&&\ex{\abs{\cP}} =k+\paren{n-k}\paren{1-\paren{1-q}p\paren{1-p}^k}^m.
\label{eqn:Thm2}
\end{eqnarray}
\end{theorem}

\begin{IEEEproof}
The proof is given in Appendix \ref{app:Proof2}.
\end{IEEEproof}

Theorem~\ref{thm:additive_noise} extends  Theorem~\ref{thm:noiseless} to the case of additive noise. Note that when $q=0$, i.e., the effect of the additive noise vanishes, we obtain the corresponding expected value derived in Theorem~\ref{thm:noiseless} for the noiseless case.

To exemplify the difference between the expected number of \ac{pd} subjects for noiseless and noisy setups, we next evaluate \eqref{eqn:Thm2} with the  parameters used in Fig.~\ref{fig:noiseless_sim} for the noiseless case. Specifically, we set $m=47,n=105$, and focus \rev{on} $k=5$ defective items. In the presence of additive noise with Bernoulli parameter $q=0.1$, we obtain that $\ex{\abs{\cP}}=98.52$, as opposed to $\ex{\abs{\cP}}=9.3$ in the noiseless case. This demonstrates the importance of adjusting the number of tests in the additive noise scenario: when we use a number of tests that corresponds to the noiseless case, \ac{dnd} on average outputs that most of the items are \ac{pd}.

\vspace{-0.3cm}
\subsection{Dilution Noise}\label{subsec:dilution}
\vspace{-0.1cm}
In the dilution noise model, a test may come negative even if defective items participate it~\cite{atia2012boolean}. This case reflects the possibility of false negatives in traditional binary \ac{gt}, which are often more severe than false positives, especially in the context of COVID-19. In particular, if a defective item is declared \ac{dnd} in Step 1 of Algorithm~\ref{alg:recovery}, then it will not be detected as infected by the algorithm.

To model the presence of dilution noise, we let each defective item be diluted with probability $u$, independently of other defective items in the test. \rev{In} order to handle the presence of dilution noise, the \ac{dnd} algorithm has to be adjusted. In the modified \ac{dnd} algorithm, a patient is declared \ac{pd} if it participates in at least $\tau$ tests that come out positive~\cite{chan2011non, chan2014non}. As a result, the number of positive tests that an item participates in is distributed according to a binomial distribution with $m$ trials. The probability to participate in a test depends on whether the item is infected or not. We denote the probability that an infected/non-infected item participates in a positive result by $\phu/\ptu$, respectively. Consequently, the number of positive tests an infected/non infected item participates in is distributed according to ${\rm Bin}\paren{m, \phu}$, ${\rm Bin}\paren{m, \ptu}$, respectively. Here, ${\rm Bin}\paren{a,b}$ is the binomial distribution with $a$ trials and probability of success $b$.

In the presence of dilution noise, we are no longer guaranteed that a defective item would always \rev{be} declared as \ac{pd}. We define $\cP_{\cD}, \cP_{\cD^c}$ as the set of all defective/non defective items that are marked \ac{pd}, respectively, i.e., $\cP_{\cD^c} \cup \cP_{\cD} = \cP$ while $\cP_{\cD^c} \cap \cP_{\cD}$ is an empty set. The extension of Theorem~\ref{thm:noiseless} to the presence of dilution noise is given in the following theorem:
\begin{theorem}\label{thm:dilution}
Let $u$ denote the dilution probability of a defective item in each test, and $\tau$ be the  threshold used by the modified \ac{dnd} algorithm. Then, the set subjects declared \ac{pd} is comprised of the distinct sets  $\cP_{\cD}, \cP_{\cD^c}$ which satisfy:
\begin{eqnarray}
&&\ex{\abs{\cP_{\cD^c}}}=\paren{n-k}\cdot\pr{{\rm Bin}\paren{m,\ptu}\geq \tau}, \nonumber \\
&&\ex{\abs{\cP_{\cD}}}=k\cdot\pr{{\rm Bin}\paren{m,\phu}\geq \tau}, \nonumber
\end{eqnarray}
and thus
\begin{align}
    \ex{\abs{\cP}}=&\paren{n-k}\cdot\pr{{\rm Bin}\paren{m,\ptu}\geq \tau} \notag \\
    &+ k\cdot\pr{{\rm Bin}\paren{m,\phu}\geq \tau}.
    \label{eqn:DilThm}
\end{align}
\end{theorem}

\begin{IEEEproof}
The proof is given in Appendix \ref{app:Proof3}.
\end{IEEEproof}
Theorem~\ref{thm:dilution} characterizes the overall number of \ac{pd} items, as done in Theorems~\ref{thm:noiseless}-\ref{thm:additive_noise} for the noiseless and additive noise setups, respectively. In addition to expressing $\ex{\abs{\cP}}$, Theorem~\ref{thm:dilution} also identifies the expected number of patients which are not defective but are identified as \ac{pd} by the \ac{dnd} algorithm.
Unlike the models used in deriving Theorems~\ref{thm:noiseless},\ref{thm:additive_noise}, under the dilution noise model we are no longer guaranteed that a defective item is declared \ac{dd}. Thus, there is a non-zero probability that some  of the $k$ infected patients are not assigned into $\cP$ by the \ac{dnd} algorithm, and thus $\ex{\abs{\cP_{\cD}}}\leq k$.

\vspace{-0.3cm}
\subsection{Uniqueness of \ac{ls} Recovery}\label{subsec:correctnext_least_squares}
\vspace{-0.1cm}
\rev{Next}, we identify sufficient conditions under which \rev{the \ac{ls} solution is unique.} Step 2 recovers the infection levels based on the output of Step 1, i.e., $\cD$. Therefore, to  provide  a lower bound on the probability that this output enables recovery of infection levels via \ac{ls} followed by discretization of the values, in the following we focus on the case where all the infected items have been identified by the first step of the algorithm.

Once the set of infected items, whose indices are denoted by $\cD$, has been identified by the \ac{gt} procedure in Step 1, the infection levels of those subjects are recovered by quantizing the \ac{ls} solution to \eqref{eqn:LS}. Notice that since we are left with $k$ subjects to resolve, it holds that $\AcD\in\cbrace{0,1}^{m\times k}$, and $m>k$. When solving ~\rev{(\ref{eqn:LS})}, we treat $\AcD$ as a real matrix rather than a matrix over the binary field. \off{It is known that a} \rev{A} unique solution exists to the \ac{ls} problem in this scenario if $\AcD$ is of full rank, namely $\rank{\AcD}=k$. When this holds, the \rev{unique solution to the \ac{ls} procedure is used to estimate viral loads. Our proposed Algorithm~\ref{alg:recovery} recovers the infection levels by quantizing these estimates, which is numerically demonstrated to yield accurate identification of the infection levels in Section~\ref{sec:sims}.} The following theorem bounds the probability that $\AcD$ is rank-deficient:

\begin{theorem}\label{thm:full_rank}
Let the number of tests satisfy $m=\paren{1+\epsilon}k\log_2\paren{n}$ for some $\epsilon>0$, and the Bernoulli parameter $p$ satisfy $p=1-2^{-1/k}$. Then the probability \rev{that the solution of~(\ref{eqn:LS}) is not unique} is bounded as:
\begin{eqnarray}
\pr{\rank{\AcD}<k}\leq 1-\paren{1-n^{-\paren{1+\epsilon}}}^{\rev{2^{k}-1}}. \label{eqn:full_rank}
\end{eqnarray}
\end{theorem}
\begin{IEEEproof}
The proof is given in Appendix~\ref{app:Proof4}.
\end{IEEEproof}

Theorem~\ref{thm:full_rank} guarantees that Step 2 of Algorithm~\ref{alg:recovery} \rev{uses a unique \ac{ls} estimate} with high probability, assuming that  Step 1 of the algorithm has successfully identified the infected items. The resulting probability depends on the number of pools $\Noutputs$ via the parameter $\epsilon>0$. 
In Section~\ref{sec:sims} we assess in simulation that the high probability of $\AcD$ having a full rank in Step 2  results in identifying the correct infection levels.


\vspace{-0.2cm}
\subsection{Complexity Analysis}\label{subsec:complexity}
\vspace{-0.1cm}
One of the key features of the proposed multi-level \ac{gt} methods is its reduced complexity compared to \ac{ml}-based \ac{gt}.
To fully analyze this aspect of Algorithm~\ref{alg:recovery}, we give the following proposition which characterizes its complexity, focusing on the implementation with \ac{ls} recovery.
\begin{proposition}
\label{pro:complexity}
Let $\epsilon=\bigo{1}$ be a design parameter. The overall complexity of Algorithm~\ref{alg:recovery}, when used with $m=\paren{1+\epsilon}k\log_2 n$, is given by
\begin{eqnarray}
\label{eqn:Complexity}
\bigo{{\paren{n + \binom{\abs{\cP}}{k}k + k^2}}k\log n}.
\end{eqnarray}
\end{proposition}
\begin{IEEEproof}
The complexity of \ac{dnd} is $\bigo{kn\log n}$~\cite{cohen2016secure}. Once \ac{pd} items are identified, the complexity of \rev{the} \ac{ml} algorithm is $\bigo{\binom{|\cP|}{k}km}$, as we have to compute a Boolean OR of $k$ vectors of size $m$, and compare it with $Q_{0,1}(\myVec{y})$.
The complexity of \ac{ls} depends on the dimensions of $\AcD$, but also on the specific implementation. Since $\AcD\in\cbrace{0,1}^{m\times k}$, \ac{ls} involves $\bigo{mk^2}$ operations~\cite{boyd2004convex}. We note that it is likely that the matrix $\AcD$ is sparse~\cite{eldar2012compressed}, and it becomes more sparse as $p$ decreases. Efficient algorithms for solving \ac{ls} are known in the literature when the matrix is sparse, e.g. LSQR~\cite{paige1982lsqr}.
Substituting  $m=\paren{1+\epsilon}k\log_2 n$ as the number of tests, we obtain \eqref{eqn:Complexity}, thus proving the proposition.
\end{IEEEproof}

The complexity expression in Proposition~\ref{pro:complexity} is \off{comprised of}\rev{includes} three terms, as the algorithm has three main components: the \ac{dnd} algorithm which identifies the subset \off{set} of \ac{pd} items $\cP$;
the \ac{ml} search in \eqref{eqn:GTML} on \off{the set} $\cP$; and the recovery of the infection levels using \ac{ls}.
The computational complexity of the \ac{ml} stage is dictated by the set  $\cP$, which is a random quantity depending on the observations. As shown in the previous subsections, the expected cardinality of the \ac{pd} set can be computed in closed form depending on the assumed noise model.

To analyze the dominant part of the algorithm in terms of complexity, we \rev{consider} the values $n=105,k=5$ that are used in Section~\ref{sec:sims}, which \rev{are} representative values when testing for COVID-19 using pooled \ac{pcr} measurements \cite{ghosh2020compressed}.
In this regime, it holds that $k^2 \ll n$, and so the third component (representing the \ac{ls} complexity) is negligible.
To quantify the complexity associated with \ac{ml}, which is dictated by the random set $\cP$, we adopt here the noiseless model for the binary measurements. In this case, when  the number of tests is set to $m=\paren{1+\epsilon}k\log_2 n$, which corresponds to the upper bound on the number of tests required for \ac{ml} decoding, we have that $\ex{\abs{\cP}}<2k$ (see Fig.~\ref{fig:noiseless_sim}). Under these settings and when replacing $\abs{\cP}$ with its expected value, the \ac{ml} decoder has to consider $\binom{2k}{k}=252$ combinations. This overhead is much lower than that required in order to apply \ac{ml} directly to the measurements, which involves a search over  $\binom{n}{k} > 9\cdot 10^7$ different combinations. As $\binom{2k}{k}$ is \rev{greater} than $n$, the dominant complexity factor corresponds to that required by the \ac{ml} decoder. This complexity term is not-prohibitive in the considered regime, and is of the same order as low complexity \ac{cs} based recovery, such as the \acl{omp} algorithm, which typically requires $\bigo{nk\log n}$ operations (after plugging in the number of tests $m$)~\cite{zhu2020efficient}.

\begin{figure}				
	\centering	
	\begin{subfigure}[b]{\columnwidth}
		\includegraphics[width=\linewidth]{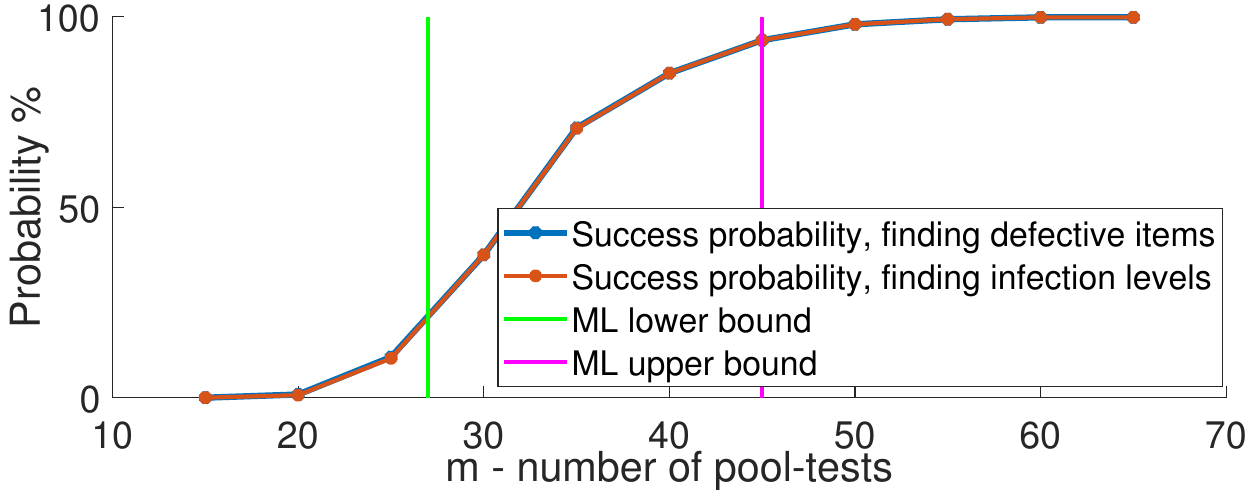}
		\caption{Success probability of multi-level \ac{gt}.}
        \label{fig:result_typical_matrix}
	\end{subfigure}
	\begin{subfigure}[b]{\columnwidth}
		\includegraphics[width=\linewidth]{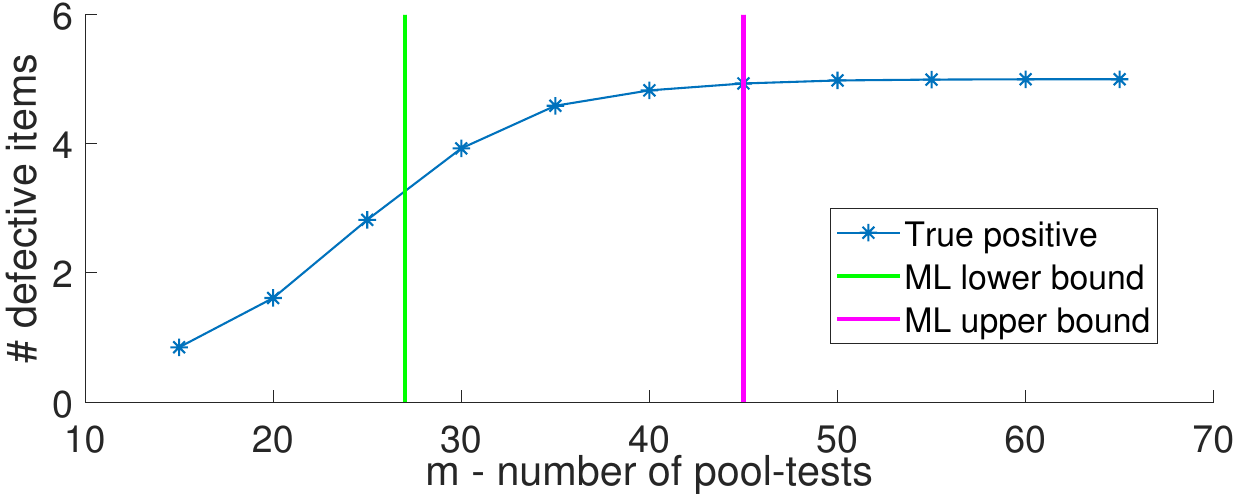}
		\caption{True positives detected in multi-level \ac{gt}.}
        \label{fig:typical_dnd_false_negative}
	\end{subfigure}
	\begin{subfigure}[b]{\columnwidth}
		\includegraphics[width=\linewidth]{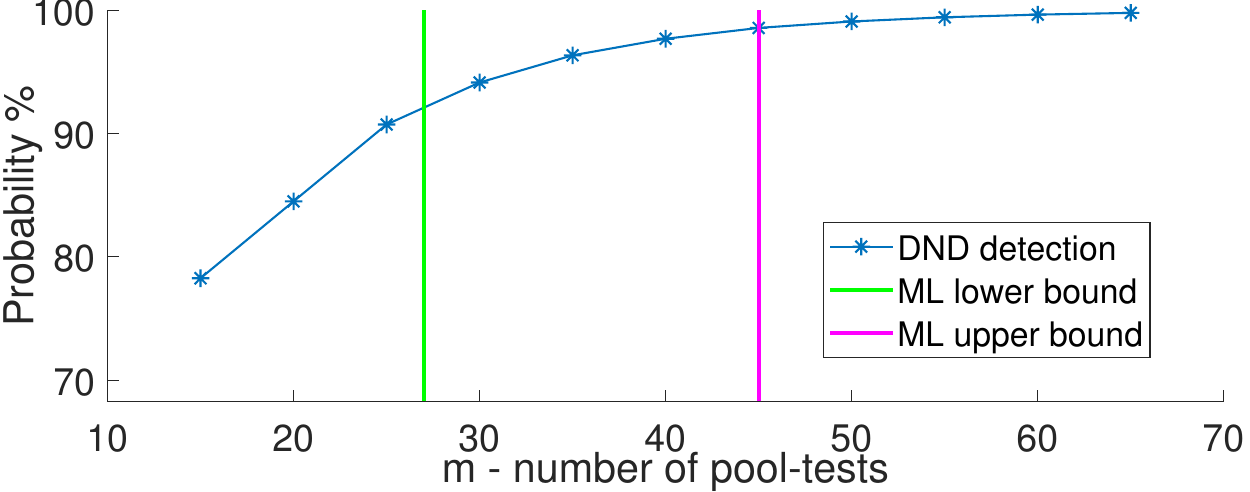}
        \caption{Detected no-defective at \ac{dnd} stage.}
        \label{fig:result_typical_probability}
	\end{subfigure}\vspace{-0.1cm}
	\caption{Performance evaluation of multi-level \ac{gt} over 200 iterations, with $\Ninputs=105$,  $\Nsparse=5$ and with $4$ infection levels.}
	\label{fig:Conf}
\end{figure}

\begin{figure*}				
	\centering	
	\begin{subfigure}[b]{.3\linewidth}
		\includegraphics[width=\linewidth, height=1.3in]{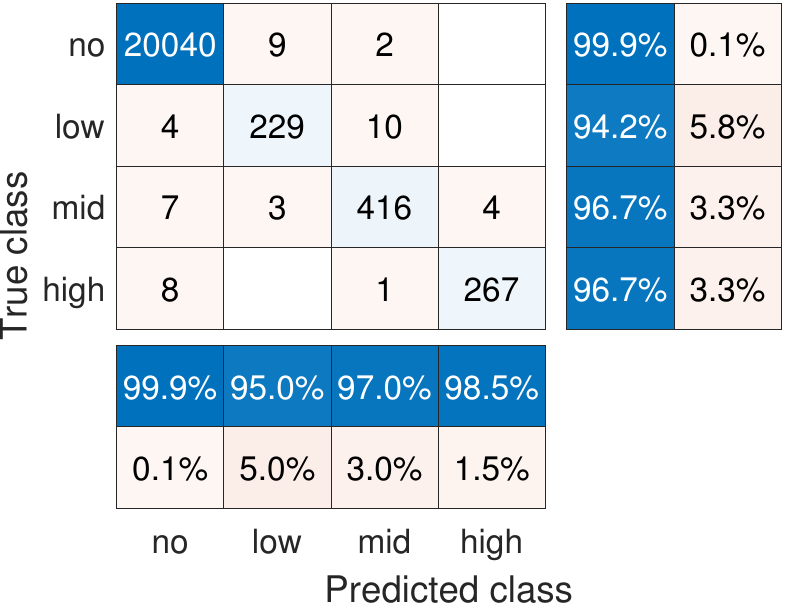}
		\caption{$n=105$, $m=45$.}\label{fig:Conf1}
	\end{subfigure}
	$\quad$
	\begin{subfigure}[b]{.3\linewidth}
		\includegraphics[width=\linewidth, height=1.3in]{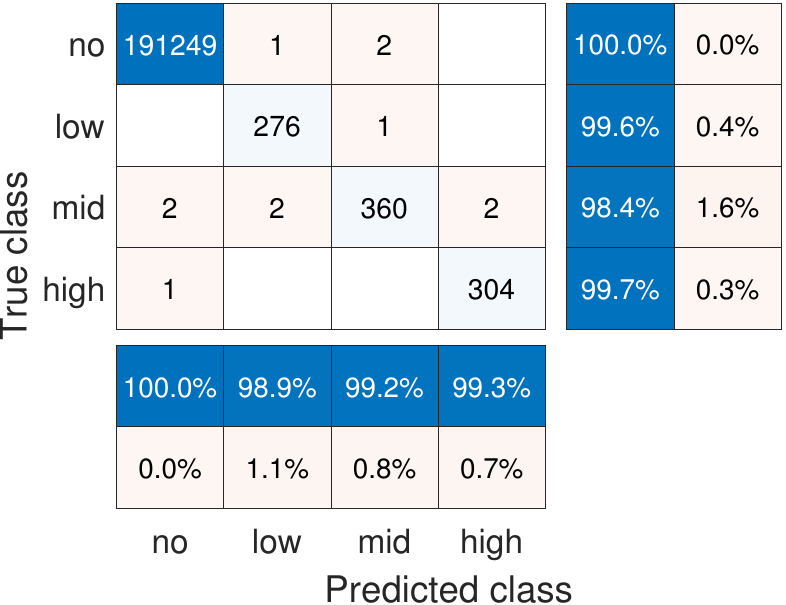}
		\caption{$n=961$, $m=70$.}\label{fig:Conf2}
	\end{subfigure}
	$\quad$
	\begin{subfigure}[b]{.3\linewidth}
		\includegraphics[width=\linewidth, height=1.3in]{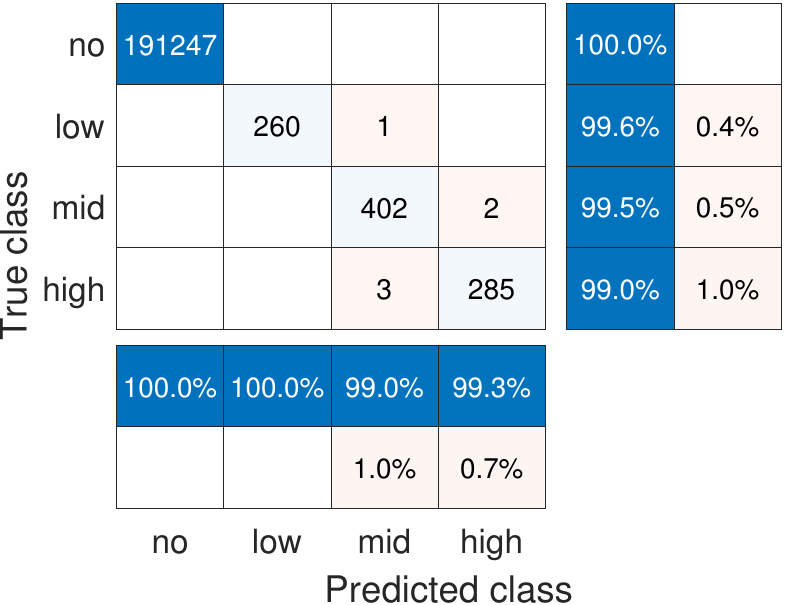}
		\caption{$n=961$, $m=93$.}
		\label{fig:Conf3}
	\end{subfigure}	
	\caption{Confusion matrices of one-shot multi-level \ac{gt} with \ac{ls} method.}
	\label{fig:ConfNew}
\end{figure*}

\begin{figure*}		
	\centering	
	\begin{subfigure}[b]{.3\linewidth}
		\includegraphics[width=\linewidth, height=1.3in]{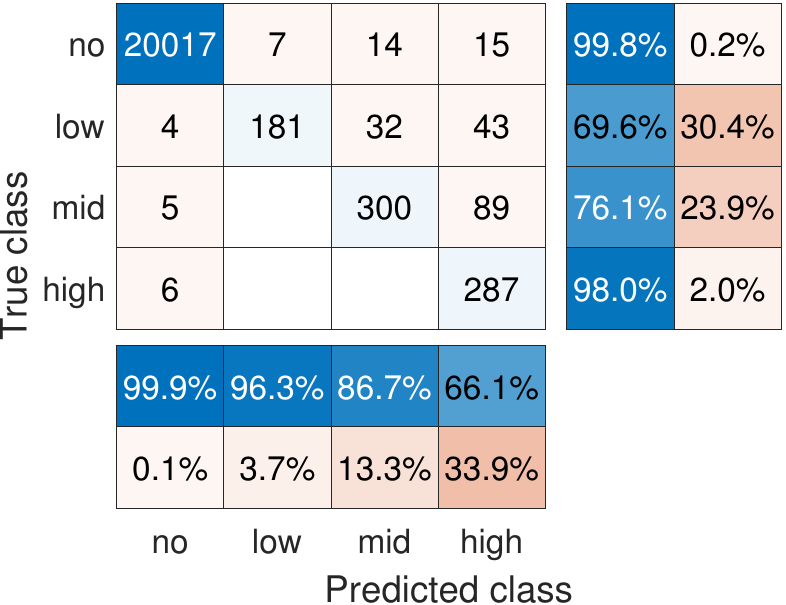}
		\caption{$n=105$, $m=45$.}\label{fig:Conf1_iterative}
	\end{subfigure}
	$\quad$
	\begin{subfigure}[b]{.3\linewidth}
		\includegraphics[width=\linewidth, height=1.3in]{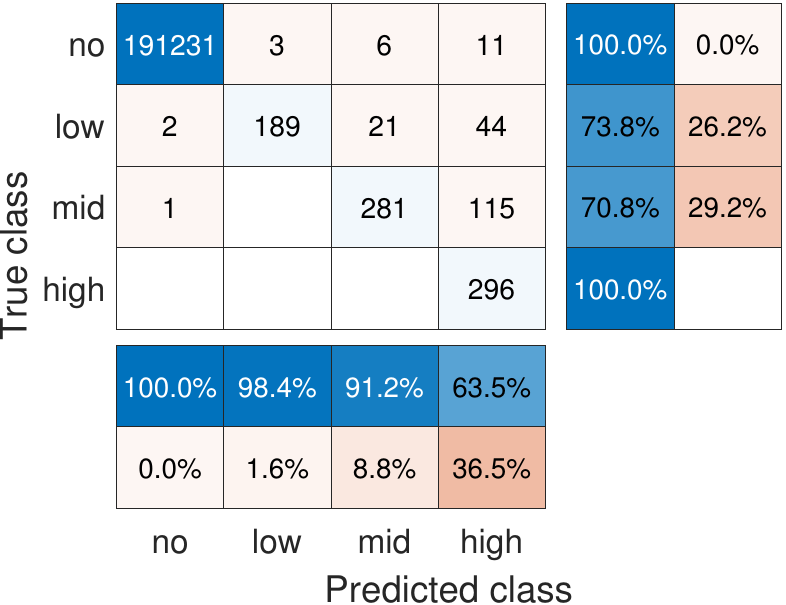}
		\caption{$n=961$, $m=70$.}\label{fig:Conf2_iterative}
	\end{subfigure}
	$\quad$
	\begin{subfigure}[b]{.3\linewidth}
		\includegraphics[width=\linewidth, height=1.3in]{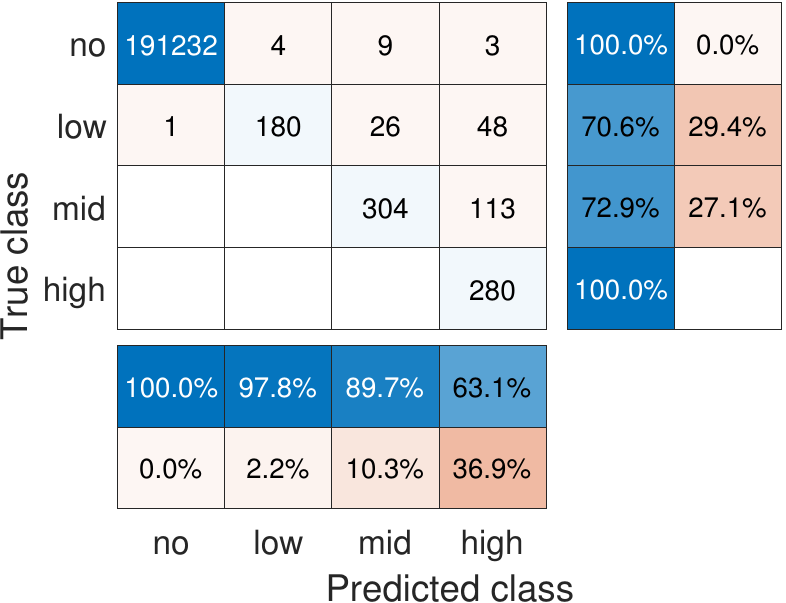}
		\caption{$n=961$, $m=93$.}
		\label{fig:Conf3_iterative}
	\end{subfigure}	
	\caption{Confusion matrices of one-shot multi-level \ac{gt} with iterative method.}
	\vspace{-0.6cm}
	\label{fig:Conf_iterative}
\end{figure*}

\vspace{-0.2cm}
\section{Numerical Evaluations}\label{sec:sims}
\vspace{-0.1cm}
In \rev{this} section we numerically evaluate the performance of the proposed multi-level \ac{gt} method recovering COVID-19 infection levels from pooled \ac{pcr} measurements.

\subsection{Experimental Setup}
\label{ssec:simSetup}
\off{To that aim, we consider the following noise model for representing the \ac{pcr} operation. i.e., the element wise mapping $f(\cdot)$ which yields $\myVec{y}=f(\myVec{z})$. In particular, the \ac{pcr} measurement operation is modeled via
\begin{equation}
   \myVec{y}_i =  \myVec{z}_i(1+q), \quad i=1,\ldots,\Noutputs,
\end{equation}
where $q\sim\cN\paren{0,\sigma^2}$.}
We consider the model used in \cite{ghosh2020compressed}  for  representing the \ac{pcr} operation, i.e., the element wise mapping $f(\cdot)$ which yields $\myVec{y}=f(\myVec{z})$. In particular, the \ac{pcr} measurement operation is modeled via
\begin{equation}
   \myVec{y}_i =  \myVec{z}_i(1+q)^{\zeta_i}, \quad i=1,\ldots,\Noutputs,
\end{equation}
where $q$ is a constant and $\{\zeta_i\}_i$ is a set of \rev{i.i.d.} zero-mean Gaussian random variables with variance $\sigma^2$.

The elements of the test matrix are chosen such that the rows and columns of $\myMat{A}$ are typical, \rev{and satisfy requirement \ref{itm:Limit} with $L=32$, by generating matrices according to a Bernoulli i.i.d. distribution until a typical matrix is obtained}. Unless stated otherwise, Algorithm~\ref{alg:recovery} implements recovery in Step 2 via \ac{ls}, and the number of patients considered is $n=105$, out of which $k=5$ are infected. The viral load of each defective subject is drawn from a uniform distribution between $\left[0,1000\right]$ as  in \cite{ghosh2020compressed}. The infection level score is based on a division of this interval into $4$ regions with thresholds $[\tau_1, \tau_2, \tau_3] = [50, 300, 700]$, namely, $[0,50)$ = no; $[50,300)$ = low (borderline); $[300,700)$ = mid; and $>700$ = high. \rev{The threshold of 300 is the recommended threshold according to the \ac{cdc} guidelines \cite{yelin2020evaluation}.} We average over 1000 Monte Carlo simulations. We choose $q = 0.95$ and $\sigma=0.01$.

The recovery performance of the proposed Algorithm~\ref{alg:recovery} is compared to upper and lower bounds on \ac{ml} recovery in \ac{gt} theory. These bounds represent the necessary and the sufficient number of test pools $\Noutputs$ one has to use in order to be able to guarantee accurate detection of which of the subjects are infected, respectively, assuming a noiseless \ac{gt} model as detailed in Subsection~\ref{subsec:Noiseless}~\cite{atia2012boolean}. The \ac{ml} lower bound is
\begin{equation}
\label{eqn:MLlb}
    \Noutputs_{\rm ML}^{\rm lb} = \log_2 \binom{\Ninputs}{\Nsparse},
\end{equation}
while the upper bound is given by
\begin{equation}
\label{eqn:MLub}
    \Noutputs_{\rm ML}^{\rm ub} = \paren{1+\epsilon}k\log_2 n,
\end{equation}
\rev{for any $\epsilon > 0$.} The bounds in \eqref{eqn:MLlb}-\eqref{eqn:MLub} guarantee detecting  the presence of infection using computationally prohibitive \ac{ml}-based \ac{gt}; we are \rev{interested not only in detecting} which patients are defective, but also to characterize their infection levels.
In addition to the \ac{ml} bounds, we also compare the performance of Algorithm~\ref{alg:recovery} to that reported in \cite{ghosh2020compressed} and \cite{shental2020efficient}.


\vspace{-0.2cm}
\subsection{Results}
\label{ssec:simResults}
\vspace{-0.1cm}
We first evaluate the success probability of Algorithm~\ref{alg:recovery} as a function of the number of pool-tests $\Noutputs$. In particular, we evaluate the algorithm using two forms of success probability: {\em finding defective items}, for which an error is declared when there is at least a single defective subject who is not detected out of a set of $\Ninputs$ patients; and {\em finding infection levels}, where an error implies that there is at least one patient whose infection level is not correctly recovered. The resulting error probabilities are depicted in Fig.~\ref{fig:result_typical_matrix}, and compared to the \ac{ml} lower/upper bound on the number of tests $\Noutputs$ computed via \eqref{eqn:MLlb}-\eqref{eqn:MLub}, respectively.
From the plot, we see that the success probability of finding the defective items and the success probability of finding the infection levels coincide. That is, whenever the defective set was recovered successfully, the correct infection levels were also estimated successfully, indicating the validity of step 2 of Algorithm~\ref{alg:recovery}. We also see that when the number of tests is at the ML upper bound, the probability of success approaches one.

\begin{figure*}
    \centering
    \includegraphics[width=1.7 \columnwidth]{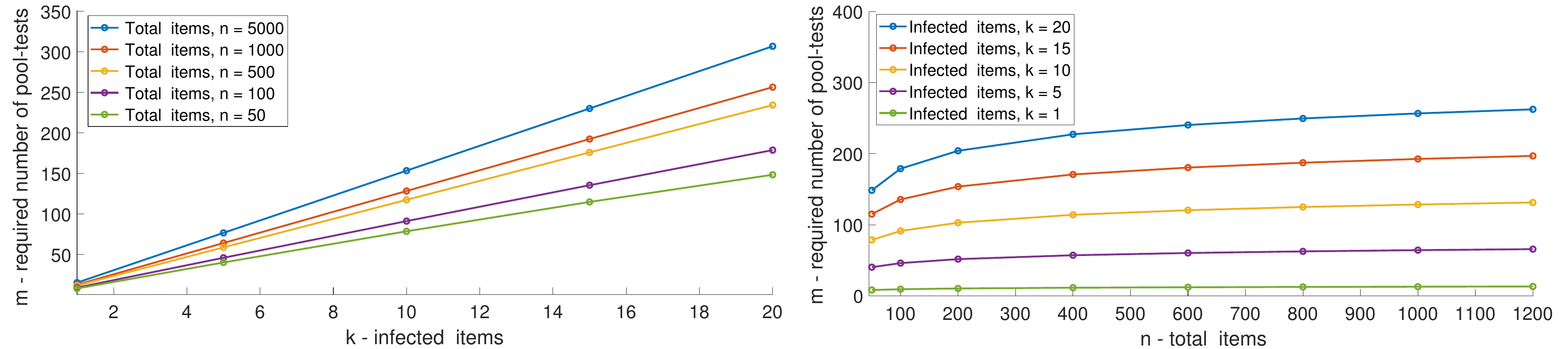}\vspace{-0.2cm}
    \caption{\rev{Required pool tests $m$ of multi-level \ac{gt} for different values of $n$ and $k$, with $\epsilon=0.02$. }}\vspace{-0.4cm}
    \label{fig:MKN}
\end{figure*}

\begin{figure}
    \centering
    \includegraphics[width=1 \columnwidth]{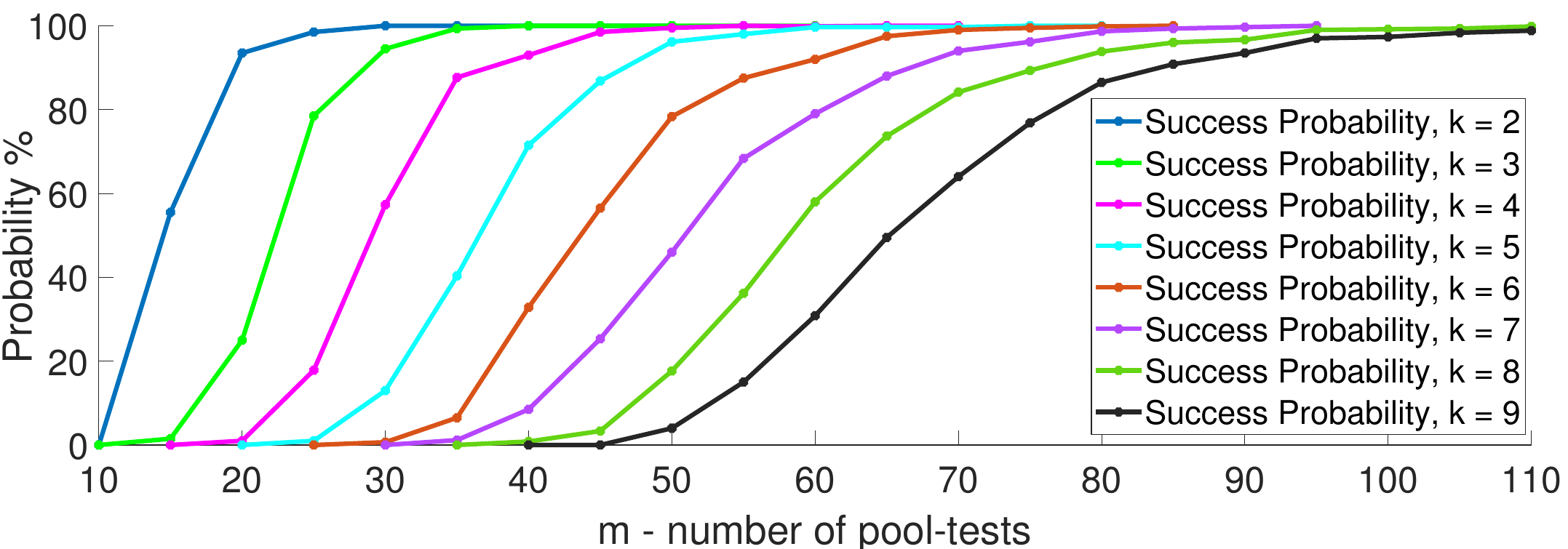}\vspace{-0.2cm}
    \caption{\rev{Success probability of multi-level \ac{gt} over $200$ iterations, with $n=200$ and $k$ varies from $2$ to $9$.}}\vspace{-0.6cm}
    \label{fig:func_k}
\end{figure}

\begin{figure}
    \centering
    \includegraphics[width=1 \columnwidth]{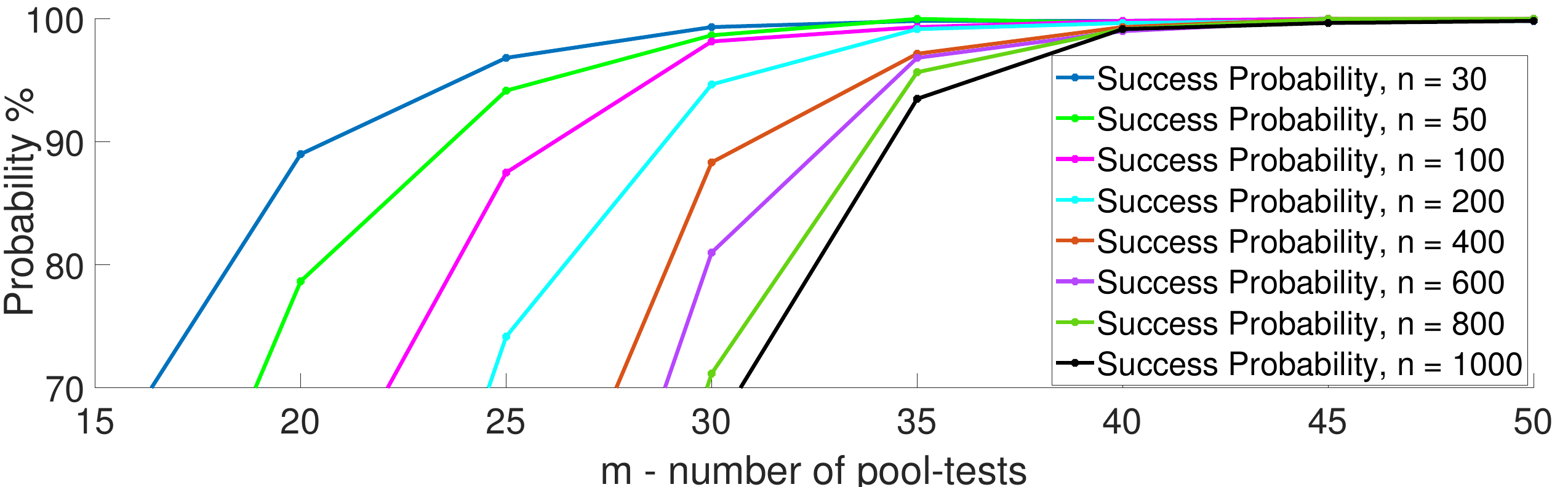}\vspace{-0.2cm}
    \caption{\rev{Success probability of multi-level \ac{gt} over $200$ iterations, with $k=3$ and $n$ varies from $30$ to $1000$.}}
    \label{fig:func_n}
\end{figure}

\off{\begin{figure}
    \centering
    \includegraphics[width=1 \columnwidth]{result_typical_code_k5_n105_2}
    \caption{Success probability of one-shot multi-level \ac{gt}.}
    \label{fig:result_typical_matrix}
\end{figure}

\begin{figure}
    \centering
    \includegraphics[width=1 \columnwidth]{typical_dnd_false_negative_k5_n105}
    \caption{True positives of one-shot multi-level \ac{gt}.}
    \label{fig:typical_dnd_false_negative}
\end{figure}

\begin{figure}
    \centering
    \includegraphics[width=1 \columnwidth]{typical_dnd_prob_k5_n105_1}
    \caption{Probability of detected non-defective at \ac{dnd} stage.}
    \label{fig:result_typical_probability}
\end{figure}
}

Fig.~\ref{fig:typical_dnd_false_negative} shows the number of true positives, i.e. the number of defective subjects that are declared defective. We observe that when  $\Noutputs$ matches the \ac{ml} upper bound, we have no false negatives with high probability. In the context of COVID-19, false negatives are the worst outcomes of a test. Fig.~\ref{fig:result_typical_probability} illustrates the probability of the detected non-defective subjects after the \ac{dnd} stage in Algorithm~\ref{alg:recovery}. This is calculated as the number of non-defective items declared by the \ac{dnd} algorithm, divided by the total number of non-defective items $n-k$. We see that when the number of tests satisfies the \ac{ml} upper bound, \ac{dnd} identifies $\approx \frac{95}{100} = 95\%$ of the subjects as non-defective, i.e., $|\cP| \approx 2k$. These will be candidates to be tested in the second stage  as \ac{pd}, demonstrating the notable complexity reduction achieved by the two-step process. \rev{Similar results are obtained when the non-negativity constraint is enforced on $\myVec{x}_{\cD}$ in the \ac{ls} formulation.}

The results in Figs. \ref{fig:result_typical_matrix}-\ref{fig:result_typical_probability} evaluate the error probability, but do not capture which types of errors are produced. We thus report in \rev{Fig.~\ref{fig:ConfNew}} the confusion matrix for the considered scenario, as well as when repeating the setup with a much larger amount of patients of $\Ninputs = 961$ \rev{(and with the same number of infected patients $k=5$)}, using merely $\Noutputs \in \{70, 93\}$ pool-tests. 
We observe in \rev{Fig.~\ref{fig:ConfNew}}  that the algorithm rarely classifies infected items as non-infected, and vice versa. For example, for $(\Ninputs, \Noutputs) = (105,45)$, only $\approx 0.1\%$ of infected items are declared non-infected, akin to the results achieved in~\cite{ghosh2020compressed} for the same parameters. \off{we note that most of the errors reported in fact correspond to identifying low-level and mid-level subjects as mid and high, respectively. Such errors are much less harmful in COVID-19 tests compared to reporting non-infected subjects as defective, which occurs only $\approx 0.1\%$ of the times for $(\Ninputs, \Noutputs) = (105,45)$ in Fig.~\ref{fig:Conf1}, which is similar to the results achieved in \cite{ghosh2020compressed} for such setups.}
This behavior is more notable when jointly testing $\Ninputs=961$ subjects in Figs. \ref{fig:Conf2}-\ref{fig:Conf3}, where for example for $\paren{\Ninputs,\Noutputs}=\paren{961,93}$ (Figure~\ref{fig:Conf2}), no false negatives are reported. Comparing these results to \cite{ghosh2020compressed}, it is noted that multi-level \ac{gt} achieves improved false positive and false negative probabilities with only $\Noutputs=70$ test-pools compared to that achieved using all \ac{cs} methods examined in \cite{ghosh2020compressed} with $\Noutputs=93$ test-pools. For instance, for $(\Ninputs, \Noutputs) = (961,93)$ \cite{ghosh2020compressed} reported false positive probabilities varying from $0.1\%$ to $0.8\%$\off{, while the corresponding probability in Fig.~\ref{fig:Conf2} is $0.0\%$}. This indicates the potential of multi-level \ac{gt} in facilitating pooled testing of large numbers of subjects.

The gains noted in Fig.~\ref{fig:Conf1} stem from the combination of \ac{gt} tools (Step 1 in Algorithm~\ref{alg:recovery}) with \ac{ls} based recovery (Step 2) which identifies the infection levels and handles the presence of noise in the \ac{dd} subjects. To quantify the gains of using \ac{ls} recovery, we repeat the same simulation where the iterative method (Option 2 for Step 2 in Algorithm~\ref{alg:recovery}) is used, reporting the confusion matrices are in Fig.~\ref{fig:Conf_iterative}. We note that here,  \ac{ls}   outperforms the iterative method. The iterative method does not attempt to denoise the measured viral loads, thus it is more susceptible to noise. This problem is mitigated by the \ac{ls} method, which inherently tries to estimate viral loads that would minimize the sum of squared errors between the measured viral loads and the estimated ones.

\rev{In Fig.~\ref{fig:MKN}, we depict the performance of multi-level \ac{gt} as a function of different numbers of infected items and total items. The results presented are for the upper bound of the required number of pool-tests, $\Noutputs_{\rm ML}^{\rm ub}$,  using ML decoder as given in \eqref{eqn:MLub}. It can be clearly seen that the number of pool-tests scales linearly with $k$ and logarithmic with $n$. In Fig.~\ref{fig:func_k}, we demonstrate the linear trend showing the error probability using Algorithm~\ref{alg:recovery} over $200$ iterations, with $n = 200$ and $k$ varied from $2$ to $9$. In Fig.~\ref{fig:func_n}, we demonstrate the logarithmic trend for $k=3$ and $n$ varies from $30$ to $1000$.}

\begin{figure}
    \centering
    \includegraphics[width=1 \columnwidth]{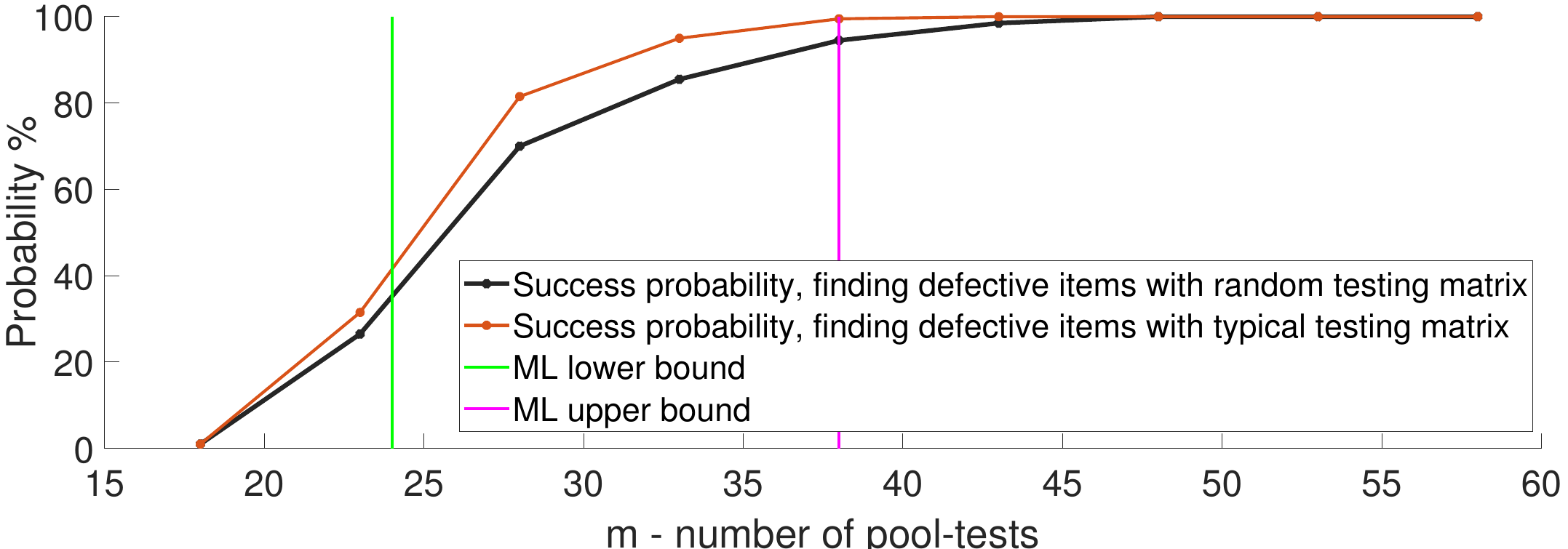}\vspace{-0.2cm}
    \caption{\rev{Success probability comparison of multi-level \ac{gt} using typical and i.i.d. testing matrices with $\paren{\Ninputs,k}=\paren{384,3}$.}}\vspace{-0.3cm}
    \label{fig:typical_vs_random_code}
\end{figure}

\rev{Next, we compare the performance achievable when using the typical and random pool-testing matrices described in Section~\ref{sec:MultiGT}. Fig.~\ref{fig:typical_vs_random_code} plots the success probability of multi-level \ac{gt} using both generation methods with $\paren{\Ninputs,k}=\paren{384,3}$. We see that when $\myMat{A}$ is typical, one can identify the infected items with about $21\%$ fewer pool-tests, and that the performance achieved with random $\myMat{A}$ also holds with typical $\myMat{A}$.}

\begin{figure}
    \centering
    \includegraphics[width= \columnwidth]{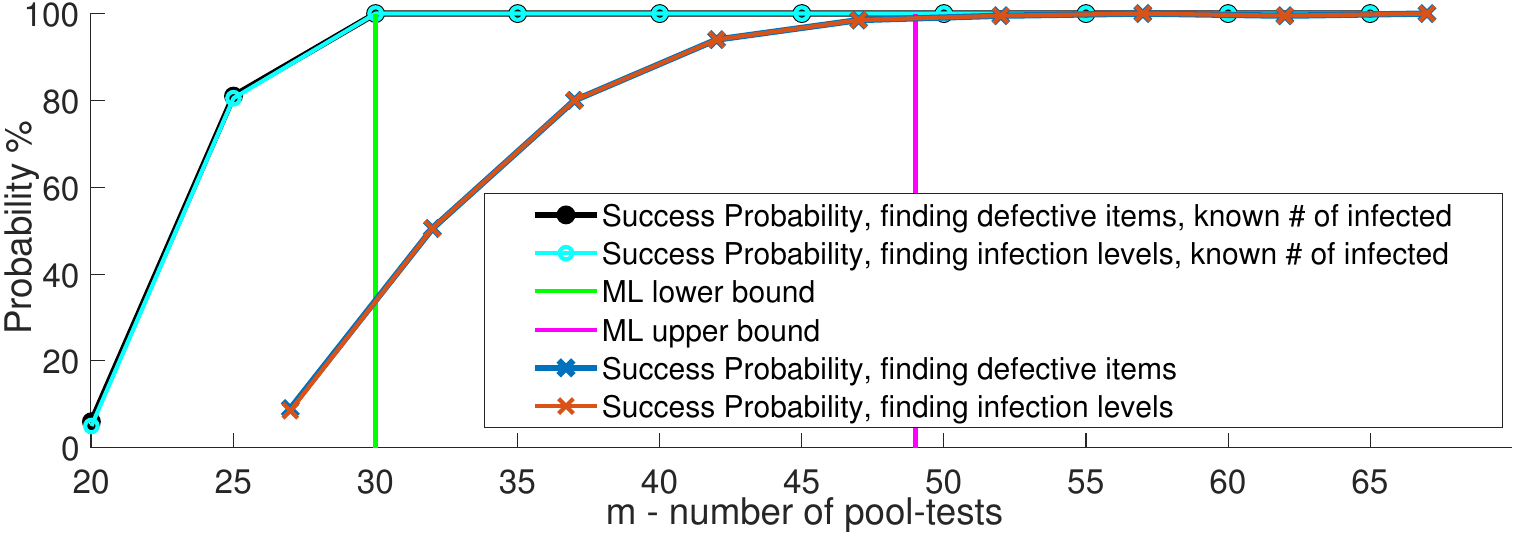}\vspace{-0.2cm}
    \caption{Success probability of multi-level \ac{gt} when the decoder can identify the number of infected subjects in a pool.}
    \label{fig:lb_sim}
\end{figure}

\rev{In} Section~\ref{sec:MultiGT} we describe how in the first \ac{gt}-based step of Algorithm~\ref{alg:recovery}, the measured pooled viral loads are represented using the mapping $\qzovy$. We use this quantity for each pooled test as an indication to whether it  includes an infected item or not; nonetheless, we do not have a similar indication regarding the number of infected items in a pooled test. We next explore how Algorithm~\ref{alg:recovery} fares when given access to such an indication on how many infected items participate in each pooled test. While not being the main focus of this paper, this scenario may be useful if we are able to output the number of infected items in each pooled test. When the number of infected items in a pooled test can be estimated in the recovery process, we \rev{may} utilize such knowledge to reduce the number of pooled tests to the known lower bound for \ac{ml} decoding of \ac{gt}~\cite{atia2012boolean}. Fig.~\ref{fig:lb_sim} plots the success probability measures as in Fig.~\ref{fig:result_typical_matrix} versus $\Noutputs$, when the number of infected items in a test is either known (in teal circles) or unknown (in orange exes) 
The number of infected items and total number of items $\paren{\Ninputs,k}=\paren{384,4}$, similar to the experiment reported in~\cite{shental2020efficient}. Fig.~\ref{fig:lb_sim} shows that like~\cite{shental2020efficient}, multi-level \ac{gt} is able to correctly identify the infected items. Unlike~\cite{shental2020efficient}, multi-level \ac{gt} is also able to recover the correct infection level of the infected items, when the number of tests is set according to the upper bound for \ac{ml} decoding. Moreover, if the number of infected items in each pooled test is revealed, multi-level \ac{gt} \rev{also achieves} the lower bound required for \ac{ml} detection, namely, it can operate reliably with the minimal necessary number of test pools dictated by \ac{gt} theory. Note that without this additional information, i.e., the number of infected items in each test, no existing solution in the literature achieves this lower bound.

\section{Conclusions}\label{sec:Conclusions}
In this work we proposed multi-level \ac{gt} for one-shot pooled COVID-19 tests. We first identified the unique characteristics and requirements of \ac{pcr}-based tests. Based on these requirements, we designed multi-level \ac{gt} to combine traditional \ac{gt} methods with one-shot operation and multi-level outputs, while implementing a preliminary \ac{dnd} detection mechanism to facilitate recovery at reduced complexity. We provided theoretical analysis of the algorithm, and identified conditions under which it operates with feasible complexity, which include the regimes of interest in the context of COVID-19. We numerically demonstrate that multi-level \ac{gt} reliably identifies the infection levels when examining  much less samples compared to the number of tested subjects.

\begin{appendix}
	\numberwithin{proposition}{subsection}
	\numberwithin{lemma}{subsection}
	\numberwithin{corollary}{subsection}
	\numberwithin{remark}{subsection}
	\numberwithin{equation}{subsection}	

	%
	\subsection{Proof of Theorem \ref{thm:noiseless}}
	\label{app:Proof1}	
	In our proof, we use $\aimu,\ai$ to denote the event that the $i$-th item is declared \ac{pd} at the $j$-th test/all of the tests, respectively. There are only two disjoint events in which a non-defective item $i$ is declared \ac{pd} in test $j$. For that to happen, either item $i$ does not participate in test $j$, or item $i$ participates in test $j$, but at least one of the defective items participate in test $j$ as well. Consequently, the only case in which item $i$ is declared \ac{dnd} at test $j$ is if it participates at test number $j$, while none of the defective items participate in that test. 

\off{\begin{figure}
    \centering
    \includegraphics[width=1 \columnwidth]{./fig/GT_example}
    \caption{Group testing example, noiseless scenario. Out of $n=5$ items, there are $k=2$ defective items, whose indices are $4,5$, i.e. $\cK=\cbrace{4,5}$. The defective items are colored in red. The total number of tests is $m=4$. The binary matrix $\myMat{A}$ is shown in the table.}
    \label{fig:gt_example}
\end{figure}}

The probability $\pr{\aimu}=1-\pr{\aimu^c} $ is calculated as:
\begin{align}
\pr{\aimu}
%
&=1-\Pr\big(\cap_{j'\in\cK}\myMat{A}\paren{j',j}=0\cap \myMat{A}\paren{j,i}=1\big)  \nonumber \\
&=1-p\paren{1-p}^k=\pt, \nonumber
\end{align}
using the fact that the elements of $\myMat{A}$ are chosen independently, and $A^c$ denotes the complement of $A$, and recall that $\cK$ denotes the set of actual defective items. Observe that $\aimu$ does not depend on $i,j$, and that
$
\pr{\ai}=\Pr\big(\cap_{j=1}^m \aimu\big) = \pt^m$,
where we use the fact that the tests are independent. Note that $\pr{\ai}$ does not depend on $i$. Let $I_i$ be the indicator function of the event that the $i$-th (non-defective) item is declared \ac{pd}, and thus, $\ex{I_i}=\pr{\ai}$. Let $D$ denote the number of non-defective items that are declared \ac{pd} by the \ac{dnd} algorithm. Then $D=\sum_{i\not\in\cK}I_i$. The expected number of non-defective items that are declared \ac{pd} by the \ac{dnd} algorithm, $\ex{D}$, is: 
\[
\ex{D}=\ex{\sum_{i\not\in\cK}I_i}=\sum_{i\not\in\cK}\ex{I_i}=\paren{n-k}\pr{\ai}.
\] Since \ac{dnd} declares all defective items as \ac{pd}, the expected number of items declared \ac{pd} is $k+\paren{n-k}\pr{\ai}$.
	
	\vspace{-0.3cm}
	\subsection{Proof of Theorem \ref{thm:additive_noise}}
	\label{app:Proof2}	
    \vspace{-0.1cm}

In the presence of additive noise, we have to adjust the calculation of $\pr{\aimu},\pr{\ai}$. The event $\aimu$ can now happen without the presence of defective items in test number $j$ with probability $q$. In other words:
\begin{align}
\pr{\aimu}
%
&=1\!-\!\Pr\big(\mathop{\cap}\limits_{j'\in\cK}\myMat{A}\paren{j,j'}\!=\!0\cap \myMat{A}\paren{j,i}\!=\!1\cap \wmu\!=\!0\big)\nonumber\\
&=1-\paren{1-q}p\paren{1-p}^k=\ptq\nonumber.
\end{align}
    \off{\begin{align}
    \pr{\aimu}&=1-\pr{\aimu^c} \nonumber \\
    &=1\!-\!\Pr\big(\mathop{\cap}\limits_{j'\in\cK}\myMat{A}\paren{j,j'}\!=\!0\cap \myMat{A}\paren{j,i}\!=\!1\cap \wmu\!=\!0\big)  \nonumber \\
    &=1-\paren{1-q}p\paren{1-p}^k=\ptq.
    \end{align}}
Akin to the proof of Theorem~\ref{thm:noiseless}, we get that $\pr{\ai}=\ptq^m$ and that $\ex{D}=\paren{n-k}\ptq^m$.
This concludes the proof of \eqref{eqn:Thm2}.
	
	\vspace{-0.3cm}
	\subsection{Proof of Theorem \ref{thm:dilution}}
	\label{app:Proof3}	
    \vspace{-0.1cm}
	Let $\mmu$ be the number of defective items that participate in test $j$, and $i$ be an index of a non-defective item. Note that $\mmu\sim {\rm Bin}\paren{k,p}$. We calculate $\pr{\aimu} = 1-\pr{\aimu^c}$ as:
\begin{align}
\pr{\aimu}
%
&=1-\sum_{j'=0}^k \pr{\mmu=j'}\pr{\aimu^c\mid\mmu=j'}  \nonumber \\
&=1- \sum_{j'=0}^k\binom{k}{j'}p^{j'}\paren{1-p}^{k-j'}pu^{j'} = \ptu\nonumber.
\end{align}
If $i$ is a defective item, the $i$-th item has to participate \rev{in the test}. \rev{The modified probability $\pr{\aimu}=1-\pr{\aimu^c}$ is}:
\begin{align}
\hspace{-0.2cm}\pr{\aimu}
%
%
&=1- \sum_{j'=0}^{k-1}\binom{k}{j'}p^{j'}\paren{1-p}^{k-j'}pu^{j'+1} = \phu\nonumber,
\end{align}
which is not a function of $i$. By plugging $u=0$, we get the noiseless case. Unlike previous cases, in dilution model, \rev{the} \ac{dnd} algorithm does not declare an item as \ac{pd} if it comes out \ac{pd} at each of the $m$ tests. Instead, item $i$ is declared \ac{pd} if it comes \ac{pd} at a sufficient amount of tests. Formally, let $\di$ be the number of times item $i$ comes out \ac{pd} by \ac{dnd} algorithm. We say that item $i$ is \ac{pd}, i.e. $\ai$ is true, if $\di\geq \tau$, where $\tau$ is a threshold. As different tests are independent, and item $i$ comes out \ac{pd} at test $j$ with probability $\pr{\aimu
}$, we conclude that $\di\sim {\rm Bin}\paren{m,\pr{\aimu}}$, as a sum of i.i.d. Bernoulli random variables. As a result, $\pr{\ai}=\pr{\di\geq\tau}$. In \rev{a} similar fashion to before, we get that $\ex{\abs{\cP_{\cD^c}}}=\paren{n-k}\cdot\pr{{\rm Bin}\paren{m,\ptu}\geq \tau}$ and $\ex{\abs{\cP_{\cD}}}=k\cdot\pr{{\rm Bin}\paren{m,\phu}\geq \tau}\rev{.}$

	\vspace{-0.3cm}
	\subsection{Proof of Theorem \ref{thm:full_rank}}
	\label{app:Proof4}	
    \vspace{-0.2cm}
We first prove the following lemma, then prove Theorem~\ref{thm:full_rank} by plugging into it $m=\paren{1+\epsilon}k\log_2\paren{n}, p=1-2^{-1/k}$.

\begin{lemma}
Let $p,k,m$ be the Bernoulli parameter of $\myMat{A}$, the number of defective items, and \rev{the} number of tests, respectively, where $k<m$ and $p\leq0.5$. Let $\AcD\in\cbrace{0,1}^{m\times k}$ be a submatrix formed by taking the $k$ columns of $\myMat{A}$ that correspond to the $k$ true defective items. Then:
\begin{eqnarray}
\pr{\rank{\AcD}=k} \geq \big(1-\paren{1-p}^{m}\big)^{\rev{2^{k}-1}}. \nonumber
\end{eqnarray}
\end{lemma}
\begin{IEEEproof}
To prove the lemma, we start with $\AcD$ that contains no columns, and gradually add $k$ columns. Let $\cbrace{\vv_j}_{j=1}^{i-1}$ be the set of all columns in $\AcD$ before adding the $i$-th column. Before adding the $i$-th column, we assume that all previously added columns $\cbrace{\vv_j}_{j=1}^{i-1}$ are linearly independent. Define $\cW_i$ as all the binary columns spanned by $\cbrace{\vv_j}_{j=1}^{i-1}$, namely:
\begin{align}
&\cW_i = \Big\{\vu\in\cbrace{0,1}^m \mid \exists \{\alpha_l\}_{l=1}^{i-1}\in\cbrace{0,1}, \vu=\sum_{j=1}^{i-1}\alpha_j\vv_j\Big\},\nonumber
\end{align}
\off{\begin{multline}
\cW_i = \nonumber \\
\cbrace{\vu\in\cbrace{0,1}^m \mid \exists \alpha_1,\ldots,\alpha_{i-1}\in\cbrace{0,1}, \vu=\sum_{j=1}^{i-1}\alpha_j\vv_j}, \nonumber
\end{multline}}
and the addition is done over the real field $\reals$. Observe that $\abs{\cW_i}\leq 2^{i-1}$, as some combinations of $\cbrace{\alpha_j}_j$ may result in columns that do not belong to $\cbrace{0,1}^m$. At each step, we add a column of length $m$ whose elements \rev{are drawn from i.i.d. Bernoulli distribution with} parameter $p$, and ask whether the added column is within the binary span of the previous columns or not, i.e. if $\vv_i\in\cW_i$. Note that $\rank{\AcD}=k$ is satisfied iff at every step, we add a column that is not in the span of the previously added columns, namely 
$\vv_1\not\in\cW_1,\ldots,\vv_k\not\in\cW_k$.
\off{\begin{eqnarray}
\rank{\AcD}=k \iff \vv_1\not\in\cW_1,\ldots,\vv_k\not\in\cW_k.\nonumber
\end{eqnarray}}
For example, in the first step, we need to add a column that is not in the span of zero previously-added columns. 
Second, we note that $\myVec{0}\in\cW_i$ for every $i$, which is the most likely column among all columns that lie in the span (since we assume that $p\leq 0.5$). The probability of choosing the zero column is given by $\pr{\vv_i=\myVec{0}}=\paren{1-p}^m$.
\off{\begin{eqnarray}
\pr{\vv_i=\myVec{0}}=\paren{1-p}^m. \nonumber
\end{eqnarray}}
As a result, the probability that the newly added column does not lie within the span of previously selected columns is bounded below by $\pr{\vv_i\not\in\cW_i} \geq \big(1-\paren{1-p}^m\big)^{2^{i-1}}$.
\off{\begin{eqnarray}
\pr{\vv_i\not\in\cW_i} \geq \big(1-\paren{1-p}^m\big)^{2^{i-1}}. \nonumber
\end{eqnarray}}

We can readily bound the probability that $\AcD$ is of full rank by repeating this argument $k$ times \rev{and getting}
$\pr{\rank{\AcD}=k}=\pr{\vv_1\not\in\cW_1,\ldots,\vv_k\not\in\cW_k}\geq \prod_{i=1}^{k}\big(1-\paren{1-p}^m\big)^{2^{i-1}} = \big(1-\paren{1-p}^m\big)^{\sum_{i=1}^{k}2^{i-1}}$, thus
\begin{align}
\pr{\rank{\AcD}=k} \geq
%
%
&=\big(1-\paren{1-p}^m\big)^{\rev{2^{k}-1}},\nonumber
\end{align}
\off{\begin{align}
&\pr{\rank{\AcD}=k}
=\pr{\vv_1\not\in\cW_1,\ldots,\vv_k\not\in\cW_k} \nonumber \\
&\geq \prod_{i=1}^{k}\big(1-\paren{1-p}^m\big)^{2^{i-1}} \nonumber
= \big(1-\paren{1-p}^m\big)^{\sum_{i=1}^{k}2^{i-1}} \nonumber \\
&=\big(1-\paren{1-p}^m\big)^{0.5\paren{2^{k}-1}},
\end{align}}
where we use the fact that the columns are chosen independently, and 
the geometric series formula.
\end{IEEEproof}
The lemma gives a lower bound on the probability of  finding a unique solution to \eqref{eqn:LS}, which is a part of Step 2, for a general selection of $p,k,m$. The proof is concluded by plugging in $m=\paren{1+\epsilon}k\log_2\paren{n}, p=1-2^{-1/k}$.
\end{appendix}

	
	\bibliographystyle{IEEEtran}
	\bibliography{IEEEabrv,refs}
	
\end{document}